\documentclass[useAMS,usenatbib,usegraphicx]{mn2e}

\usepackage{aas_macros}

\usepackage{multirow}

\usepackage{amsmath}

\title[Observing star-forming filaments]{Modelling the chemistry of star-forming filaments -- II. Testing filament characteristics with synthetic observations}
  \author[D. Seifried et al.]
  {D.~Seifried,$^{1}$\thanks{seifried@ph1.uni-koeln.de} \'A.~S\'anchez-Monge,$^{1}$ S. Suri,$^{1}$, S. Walch$^{1}$ \\
  $^1$I. Physikalisches Institut, Universit\"at zu K\"oln, Z\"ulpicher Str. 77, 50937 K\"oln, Germany}

\date{Released 2017}

\pagerange{\pageref{firstpage}--\pageref{lastpage}} \pubyear{2015}

\begin{document}

\label{firstpage}

\maketitle

\begin{abstract}
We present synthetic continuum and $^{13}$CO and C$^{18}$O line emission observations of dense and cold filaments. The filaments are dynamically evolved using 3D-MHD simulations that include one of the largest on-the-fly chemical networks used to date, which models the detailed evolution of H$_2$ and CO. We investigate the reliability of observable properties, in particular filament mass and width, under different simulation conditions like magnetic field orientation and cosmic ray ionisation rate. We find that filament widths of $\sim$0.1 pc can be probed with both line and continuum emission observations with a high accuracy (deviations $\leq$ 20\%). However, the width of more narrow filaments can be significantly overestimated by up to a factor of a few. Masses obtained via the dust emission are accurate within a few percent whereas the masses inferred from molecular line emission observations deviate from the actual mass by up to a factor of 10 and show large differences for different $J$ transitions. The inaccurate estimate of filament masses and widths of narrow filaments using molecular line observations can be attributed to (i) the non-isothermal state of the filaments, (ii) optical depth effects, and (iii) the subthermally excited state of CO, while inclination effects and opacity correction only influence the obtained masses and widths by less than 50\%. Both, mass and width estimates, can be improved by using two isotopes to correct for the optical depth. Since gas and dust temperature generally differ (by up to 25 K), the filaments appear more gravitationally unstable if the (too low) dust temperature is used for the stability analysis.
\end{abstract}


\section{Introduction}

Filaments seem to play a prominent role in the process of star formation, a result which was re-emphasised with the advent of \textit{Herschel} \citep{Andre10}. These observations suggested a rather uniform filament width of about 0.1 pc \citep[e.g.][]{Arzoumanian11,Peretto12,Palmeirim13}. However, this universal width is questioned by an increasing number of observations, which show that filaments cover a wide range of widths over about two orders of magnitude ranging from 0.01 to 1 pc \citep{Juvela12a,Malinen12,Panopoulou14,Panopoulou17,Sanchez14}, Moreover, \citet{Hacar13} show that even a single filament can consist of a bundle of individual fibres.

One reason for this wide range of widths might be different environmental conditions under which filaments form. It is known that the cosmic ray ionisation rate in our Galaxy varies from 10$^{-17}$ to 10$^{-15}$ s$^{-1}$ \citep{Vastel06,Caselli98,Ceccarelli11}. Furthermore, also the magnetic field structure inside the filament can change from either a perpendicular to a parallel orientation with respect to the main filament axis \citep{Chapman11,Sugitani11,Li13,Palmeirim13,Planck16,Pillai14}. As shown in \citet{Seifried15}, the field structure indeed significantly affects the resulting width of the filament: Whereas for the perpendicular case rather narrow filaments with widths smaller than 0.1 pc form, for the parallel case the magnetic pressure acts to stabilize the filaments against radial collapse resulting in widths of around 0.1 pc.

Furthermore, filaments are observed with different techniques and different wavelengths. Both, observations from line emission \citep[e.g.][]{Hernandez11,Arzoumanian13,Busquet13,Hacar13,Panopoulou14,Sanchez14,Friesen16,Kainulainen16a,Kainulainen16b} as well as dust emission \citep[e.g.][]{Andre10,Konyves10,Arzoumanian11,Peretto12,Kainulainen13,Palmeirim13} are available now, which might lead to somewhat different filament properties even if the same object would be observed.

All these aspects give rise to the question how reliable observed properties like the mass or the width of filaments are, which may have formed under different environmental conditions or are observed with different techniques. It is thus timely to investigate this question by means of numerical simulations combined with the use of radiative transfer calculations.

In this context we emphasise that recent numerical results also reproduce the rather large range of filament widths found by observations \citep{Smith14,Moeckel15,Seifried15}. In order to make predictions for actual observations, the subsequent radiative transfer calculations require accurate chemical abundances and/or dust temperatures. However, many numerical simulations still lack the usage of a chemical network in order to calculate the chemical abundances along with the hydrodynamical quantities, which then requires a chemical post-processing step. The applicability of such a chemical postprocessing, however, is questionable since the chemistry is not necessarily in equilibrium \citep[e.g][]{Glover07,Glover12}. Magnetohydrodynamical (MHD) simulations including such chemical networks are still limited to date but nonetheless become more and more feasible \citep[e.g][]{Dobbs08b,Inoue12,Clark13,Pettitt14,Smith14,Szucs14,Walch15,Hocuk16}. In a recent work we have performed simulations of star-forming filaments, in which we incorporated one of the \textit{largest} chemical networks used to date on-the-fly in 3D-MHD simulations \citep{Seifried16}. These simulations investigate the dynamical and chemical evolution of filaments including the formation of H$_2$ and CO as well as the most significant cooling and heating processes. Moreover, we have included a simplified radiative transfer in order to model the attenuation of the interstellar radiation field. With these techniques at hand we can, for the first time, simulate reliable chemical abundances, as well as the temperatures of both gas and dust.

In this paper we will use the simulations of \citet{Seifried16} to produce synthetic observations of both line and dust emission in order to test the reliability of observable properties like the widths and masses of star-forming filaments. The paper is organised as follows: In the Sections~\ref{sec:IC} and~\ref{sec:synobs} we briefly describe the simulations and numerical methods used to produce the synthetic observations. Next, in Section~\ref{sec:results} we present the results of our analysis concerning filaments masses and widths. In Section~\ref{sec:LOS} we discuss various line-of-sight effects which could complicate the analysis. Finally, we assess the reliability of our observed properties in Section~\ref{sec:reliability} and summarise our results in Section~\ref{sec:conclusion}.

\section{Overview of the simulations}
\label{sec:IC}

The synthetic observations discussed here are based on the simulations presented in \citet{Seifried16}. Here we briefly summarize the main points. The simulations are performed with the astrophysical code FLASH 4.2 \citep{Fryxell00,Dubey08}. We solve the equations of ideal MHD using a maximum spatial resolution of 40.3 AU. The Poisson equation for gravity is solved using a multipole method based on a Barnes-Hut tree and isolated boundary conditions (W\"unsch et al, in prep.).

The simulated filaments have an initial width of 0.1 pc following a Plummer-like density profile along the radial direction. The filaments have a length of 1.6 pc and a mass per unit-length of 75 M$_{\sun}$/pc. This corresponds to about three times the critical mass per unit length \citep{Ostriker64} of
\begin{equation}
   (M/L)_\rmn{crit} = \frac{2 c_\rmn{s}^2}{G} \, ,
 \label{eq:crit}
\end{equation}
which in our case for the initial temperature of $\sim$ 15 K is about 25 M$_{\sun}$ pc$^{-1}$. For this reason the filaments are unstable against collapse along the radial direction and subject to subsequent fragmentation. The initial central density of the filaments is $3\times10^{-19}$ g cm$^{-3}$.

We take the initial magnetic field strength in the centre of the filament to be 40 $\mu$G, in agreement with recent observations \citep[e.g.][]{Sugitani11}. The field is oriented either perpendicular or parallel to the filament. In the first case it is uniform in strength, for the latter case it declines outwards proportional to $\rho^{0.5}$. In addition, we initially superimpose a transonic turbulent velocity field \citep[for more details see][]{Seifried15}.

The strength of the interstellar radiation field (ISRF) and the cosmic ray ionisation rate (CRIR) influence the reaction rates as well as the heating of dust, which is taken into account by the chemical network (see Section~\ref{sec:chem}). The ISRF and the CRIR can vary locally in our Galaxy \citep[e.g.][]{Vastel06,Caselli98,Ceccarelli11}. For this reason we performed runs with a CRIR of $1.3 \times 10^{-17}$ s$^{-1}$, $1 \times 10^{-16}$ s$^{-1}$, and $1 \times 10^{-15}$ s$^{-1}$ and an ISRF corresponding to 1.7 times the Habing flux, i.e. $G_0$ = 1.7 \citep{Draine78}. In addition we run simulations with an increased ISRF ($G_0$ = 8.5) keeping the CRIR fixed at $1 \times 10^{-16}$ s$^{-1}$. Hence, including the two possible magnetic field configurations, in total we have a set of 2 $\times (3 + 1)$ = 8 simulations with different (initial) conditions (see Table~\ref{tab:models}).
\begin{table}
 \caption{Overview of the simulations showing the used CRIR, the ISRF in values of $G_0$ \citep{Draine78}, the magnetic field configuration, and the FWHM of the filament at $t$ = 300 kyr.}
 \begin{tabular}{ccccc}
  \hline
  CRIR & G$_0$ & B-field & FWHM at $t$ = 300 kyr\\ \relax
  [$10^{-16}$ s$^{-1}$]&&& [pc] \\
  \hline
  0.13 & 1.7 & parallel & 0.034$^{a}$\\
  0.13 & 1.7 & perpendicular & 0.0058$^{a}$\\
  1 & 1.7 & parallel & 0.064\\
  1 & 1.7 & perpendicular & 0.0059$^{a}$\\
  10 & 1.7 & parallel & 0.21\\
  10 & 1.7 & perpendicular & 0.17\\
  \hline
  1 & 8.5 & parallel & 0.068\\
  1 & 8.5 & perpendicular & 0.0088\\
  \hline
 \end{tabular}
 \\
 $^a$ denoted as condensed filaments in the paper
\label{tab:models}
\end{table}

\subsection{The chemical network}
\label{sec:chem}

In order to model the chemical evolution of the gas, we use the KROME package \citep{Grassi14}. The network used comes along with KROME and is called \texttt{react$\_$COthin}, which, to a large extend, is based on the network listed in \citet{Glover10}.
It contains 37 species and 287 reactions including different forms of hydrogen, carbon, and oxygen bearing species like H$^+$, H, H$_2$, C$^+$, C, and CO, along with more complex species like e.g. HCO$^+$, H$_2$O, or H$_3^+$. In particular, the network allows for a very detailed description of the formation of CO and H$_2$ including the formation of H$_2$ on dust.

We assume that all elements heavier than He are depleted with respect to their cosmic abundances using typical values given by \citet{Flower05} for dense molecular gas, thus accounting for the freeze-out of CO in a simplified manner. Initially all elements are in their atomic form. We note that before we start the \textit{hydrodynamical} evolution of the filaments, we initially evolve the chemistry for 500 kyr in order to obtain a reasonable chemical composition (see Section~\ref{sec:overview}). In the centre of the filaments, this results in a typical H$_2$ number density of about 6.4 $\times$ 10$^4$ cm$^{-3}$, atomic hydrogen number densities of a few times 1 cm$^{-3}$, and CO number densities around 10 cm$^{-3}$.

We calculate the attenuation of the ISRF as well as the self-shielding factors for H$_2$ and CO photodissociation \citep[][see also sections 2.2.1 and 2.2.2 in \citet{Walch15} for a more detailed technical description]{Glover10} self-consistently during the simulations using the TreeCol algorithm \citep[][W\"unsch et al., in prep.]{Clark12}. The attenuation of the ISRF also affects the photoelectric heating due to dust particles as well as the dust temperature itself. Furthermore, we use self-consistent cooling and heating routines based on the calculated chemical abundances. For more details on the chemical network and cooling routines we refer the readers to the original publication \citep[][in particular the Online Material]{Seifried16}.

\subsection{Simulation results}
\label{sec:overview}

\begin{figure}
 \includegraphics[width=\linewidth]{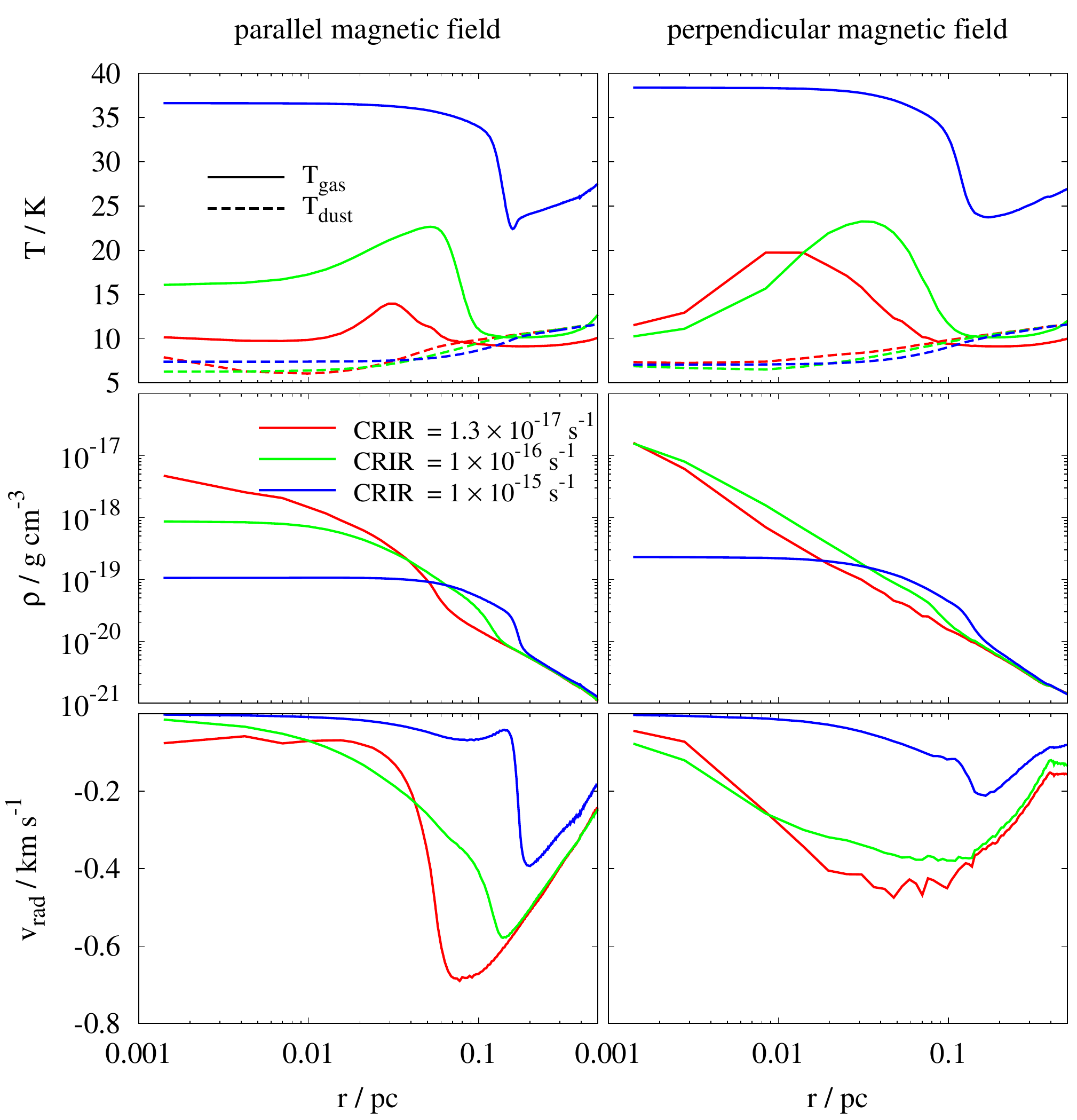}
 \caption{Radial profile of the gas (solid lines) and dust temperature (dashed lines, top row), the gas density (middle row), and the radial velocity (bottom row) at $t$ = 300 kyr averaged along the axis of the filaments with a parallel (left) and perpendicular magnetic field (right), $G_0$ = 1.7, and three different CRIRs. The gas temperature increases with increasing CRIR. In addition, the gas is heated due to an accretion shock occurring around 0.01 -- 0.1 pc, which is recognisable in an increase in both the density and the radial velocity. The dust temperature is markedly different and decreases towards the centre following a polytropic equation of state.}
 \label{fig:temp}
\end{figure}

Here, we briefly summarize the main outcomes of the simulations. Before the hydrodynamical evolution begins, the chemistry is evolved for 500 kyr. Thus, the simulations start from a rough chemical equilibrium. We obtain a gradual transition from H to H$_2$ and from C$^+$ over C to CO towards the centre of the filaments. The particular density distributions depend on the value of the ISRF and CRIR. Moreover, we find that the dust temperature $T_\rmn{dust}$ decreases towards the centre of the filament following a polytropic relation  \mbox{$T_\rmn{dust}$ $\propto$ $\rho^{\gamma-1}$} with  \mbox{$\gamma$ $\simeq$ 0.9 -- 0.95}, in agreement with recent observations \citep{Arzoumanian11,Palmeirim13,Li14}. Interestingly, the simulations show that dust and gas temperature $T_\rmn{gas}$ are markedly different even at densities of $n$ $\sim$ 10$^5$ cm$^{-3}$ (see also Section~\ref{sec:caveats}). In the top row of Fig.~\ref{fig:temp} we show the radial profiles of $T_\rmn{dust}$ and $T_\rmn{gas}$ averaged along the axis of the filaments at \mbox{$t$ = 300 kyr}\footnote{Note that throughout the paper the times used refer to the point from where we start the hydrodynamical evolution.} for the three different CRIRs considered here and \mbox{$G_0$ = 1.7} for runs with a parallel (left panel) and perpendicular (right panel) magnetic field configuration \citep[see also Fig. 3 in][]{Seifried16}. While $T_\rmn{dust}$ is rather unaffected by the CRIR with values below 12 K, $T_\rmn{gas}$ clearly increases with increasing CRIR by up to 35 K. Moreover, the profiles for $T_\rmn{gas}$ show a significant rise at a distance of about 0.01 pc to 0.1 pc from the centre depending on the actual value of the CRIR. Investigating the dynamics of the filament, we can attribute this temperature increase to an accretion shock where the infalling gas is decelerated and the density experiences a significant increase (middle and bottom rows of Fig.~\ref{fig:temp}).

The configuration of the magnetic field has a strong impact on the filament evolution \citep{Seifried15}. Supercritical filaments ($M/L \geq 3 (M/L)_\rmn{crit}$), which have either no or a perpendicular magnetic field, typically reveal a full width at half maximum\footnote{We note that for a Plummer-like density profile \mbox{$\rho(r) \propto 1/(1+(r/R_\rmn{flat})^2)^{p/2}$} as used in e.g. \citet{Andre10} and \citet{Seifried15}, the FWHM is obtained from the characteristic radius $R_\rmn{flat}$ via FWHM = $2 R_\rmn{flat} \sqrt{2^{2/p} - 1}$.}  (FWHM) which is clearly smaller than 0.1 pc. On the other hand, we find that a magnetic field that is parallel to the major axis can stabilize the filament against radial collapse resulting in widths of $\sim$ 0.1 pc. This is also reflected in the fragmentation properties. While parallel magnetic fields can suppress the formation of protostars, for the perpendicular field case the filaments show a high degree of fragmentation.

\section{Synthetic observations}
\label{sec:synobs}

\subsection{Radiative transfer}

We use the freely available radiative transfer code RADMC-3D \citep{Dullemond12} to produce synthetic continuum and line emission maps of our simulated filaments with a resolution of 0.0016 pc. We also checked the convergence of our synthetic emission maps using a two times lower resolution. We find that they are converged with relative deviations of the total flux of a few times 10$^{-3}$. For the continuum we calculate the emission at wavelengths of 70, 160, 250, 350, 500, 850, 1300, and 2600 $\mu$m. We use the opacities of \citet{Ossenkopf94} for grains with thin ice mantles coagulated at a particle density of 10$^5$ cm$^{-3}$, which corresponds to the typical densities found in our filaments. The dust temperature is provided directly by the MHD simulations.

For the line emission we consider the transitions $^{13}$CO, $J$ = 1 -- 0, $^{13}$CO, $J$ = 2 -- 1, C$^{18}$O, $J$ = 1 -- 0, and C$^{18}$O, $J$ = 2 -- 1. Since the chemistry network used in the simulations does only follow $^{12}$CO but not its isotopes $^{13}$CO and C$^{18}$O, we scale down the abundance of $^{12}$CO by 69 and 557 to obtain the abundances of $^{13}$CO and C$^{18}$O, respectively \citep{Wilson99}.

We use the Large Velocity Gradient (LVG) method to calculate the level population as well as the resulting intensity of the various line transitions. The molecular data, e.g. the Einstein coefficients, are taken from the Leiden Atomic and Molecular database \citep{Schoier05}. The line emission maps produced with RADMC-3D cover a velocity range from -1.5 km s$^{-1}$ to 1.5 km s$^{-1}$, which guarantees that all emission is captured properly (see also Section~\ref{sec:subtherm}). The channel width is 50 m s$^{-1}$, which results in 61 channels. This high spectral resolution allows us to accurately track the dynamics of the filaments. Except for Section~\ref{sec:incl}, the viewing angle is chosen such that the filaments are seen edge-on in a way that -- for the case of a perpendicular magnetic field -- the initial field lines are oriented perpendicular to the line of sight (LOS).

\subsection{Column density estimate}

In this work we only consider the transitions for $^{13}$CO and C$^{18}$O since $^{12}$CO is optically thick and thus no appropriate tracer of the structure and dynamics of the filaments \citep[e.g.][]{Bally86,Bally87}. We calculate the CO column density\footnote{Note that throughout the paper ``CO'' without any nucleon number refers to both $^{13}$CO and C$^{18}$O.} for each channel with a width d$v$ by
\begin{equation}
 N_\rmn{CO}(v) = \frac{8 \pi \nu^3}{c^3} \frac{1}{A} \, f(T_\rmn{ex}) \, \frac{Q}{g_\rmn{u}} \, \rmn{exp}\left(\frac{E_\rmn{u}}{T_\rmn{ex}}\right) \, \tau \rmn{d}v \, ,
 \label{eq:NCO}
\end{equation}
where $A$ and $\nu$ are the Einstein coefficient and frequency of the considered transition, $g_\rmn{u}$ and $E_\rmn{u}$ the degeneracy and energy (in Kelvin) of its upper state\footnote{Note that since we calculate the total column density, the degeneracy of the lower state $g_\rmn{l}$ cancels out.}, and $h$ and $k_\rmn{B}$ the Planck and the Boltzmann constant. The partition function $Q$ is evaluated at the excitation temperature $T_\rmn{ex}$ (which we have to assume, see Section~\ref{sec:extemp}). The function $f$ is given by
\begin{equation}
 f(T_\rmn{ex}) =  \frac{1}{\rmn{exp}\left(\frac{h \nu}{k_\rmn{B} T_\rmn{ex}}\right)-1}
\end{equation}
The optical depth $\tau$ for each channel is determined solving the equation
\begin{equation}
 T_\rmn{B} = \frac{h \nu}{k_\rmn{B}} \left( f(T_\rmn{ex}) - f(T_\rmn{bg}) \right) \left(1 - e^{-\tau} \right) \, ,
\label{eq:tau}
\end{equation}
where $T_\rmn{B}$ is the brightness temperature obtained from the emission map and $T_\rmn{bg}$ = 2.73 K the background temperature \citep[see e.g.][for an application of this approach in actual observations of filaments]{Hernandez11}. We note that for the calculation of $T_\rmn{B}$ we have subtracted the background emission from the observed intensity.

The total CO column density in each pixel is then obtained by summing over all velocity channels.
This can be converted into an H$_2$ column density assuming a fixed relative abundance of 10$^{-4}$/69 = $1.45 \times 10^{-6}$ and 10$^{-4}$/557 = $1.80 \times 10^{-7}$ for $^{13}$CO and C$^{18}$O, respectively, as commonly used in observational works. Hence, we obtain the total mass in each pixel via
\begin{equation}
m_\rmn{pixel} = \left .
\begin{array}{l@{}l}
 N_\rmn{^{13}CO} \times 69&\\
 N_\rmn{C^{18}O} \times 557&
\end{array}
\right \rbrace \times 10^4 \times 2.8 \, m_\rmn{p} \times A_\rmn{pixel} \, ,
\label{eq:mass}
\end{equation}
where $m_\rmn{p}$ is the proton mass, and $A_\rmn{pixel}$ the physical area of the pixel. The factor 2.8 takes into account the presence of helium.\footnote{We note that for the CO-H$_2$ conversion factor we here have adopted a fiducial value of 10$^4$ typically used in observations. This is not entirely consistent with the fractional abundance for carbon of 8.27 $\times$ 10$^{-5}$ used in the simulations \citep{Seifried16}. However, since the CO-H$_2$ conversion factor is not constant in our simulation domain due to dissociation in the outer parts \citep[see also][]{Szucs16}, the actual factor derived from the simulations directly is in the majority of the simulations indeed very close to 10$^4$. Moreover, since an observer would not have this information, we here apply the commonly used value of 10$^4$.}

From the continuum emission observations we determine the ``observed'' dust temperature $T_\rmn{dust}$ as well as the gas column density $\Sigma$ in the same manner as observers do (i.e. not those values which we obtain from the simulation data directly, see Section~\ref{sec:results}). For this purpose, we fit a modified blackbody spectrum
\begin{equation}
 I(\nu) = B(\nu,T_\rmn{dust}) \left(1 - \rmn{exp}\left[-0.1 \left( \frac{\nu}{1000 \rmn{GHz}}\right)^{\beta}  \times \frac{\Sigma}{\rmn{g \, cm^{-2}}}\right]\right)
 \label{eq:BB}
\end{equation}
to the spectral energy distribution $I(\nu)$ of each pixel
using all wavelengths from 70 $\mu$m up to 2.6 mm.
Here, $B(\nu,T_\rmn{dust})$ is the Planck spectrum, where $\nu$ denotes the frequency. Furthermore, we have used a dust opacity law of $\kappa_{\nu} = 0.1 \times ( \nu/1000 \rmn{GHz})^{\beta}$ with a constant value of $\beta = 2$ in agreement with the approach in a large number of actual observations \citep[e.g.][]{Hildebrand83,Andre10,Konyves10, Arzoumanian11,Juvela12a,Malinen12,Peretto12,Palmeirim13}.

\section{Results}
\label{sec:results}

Using the synthetic line emission maps, we can infer column densities for $^{13}$CO and C$^{18}$O, which we can convert into total mass column densities, in the same fashion as an observer would do. From the continuum emission maps we determine the ``observed'' dust temperature as well as the total mass column density (which can of course differ from that obtained from the line emission maps). Given the synthetic total column density maps, we use these to derive observable properties like mass and width of the filaments. We note that we also produced line emission maps of C$^{17}$O, which is even less abundant than C$^{18}$O (fractional abundance of about 1/1500), in order to test the reliability of our results in the optically thin limit. In general, we find qualitatively very similar results as for C$^{18}$O, which is why we do not discuss it further here.

\begin{figure}
  \includegraphics[width=\linewidth]{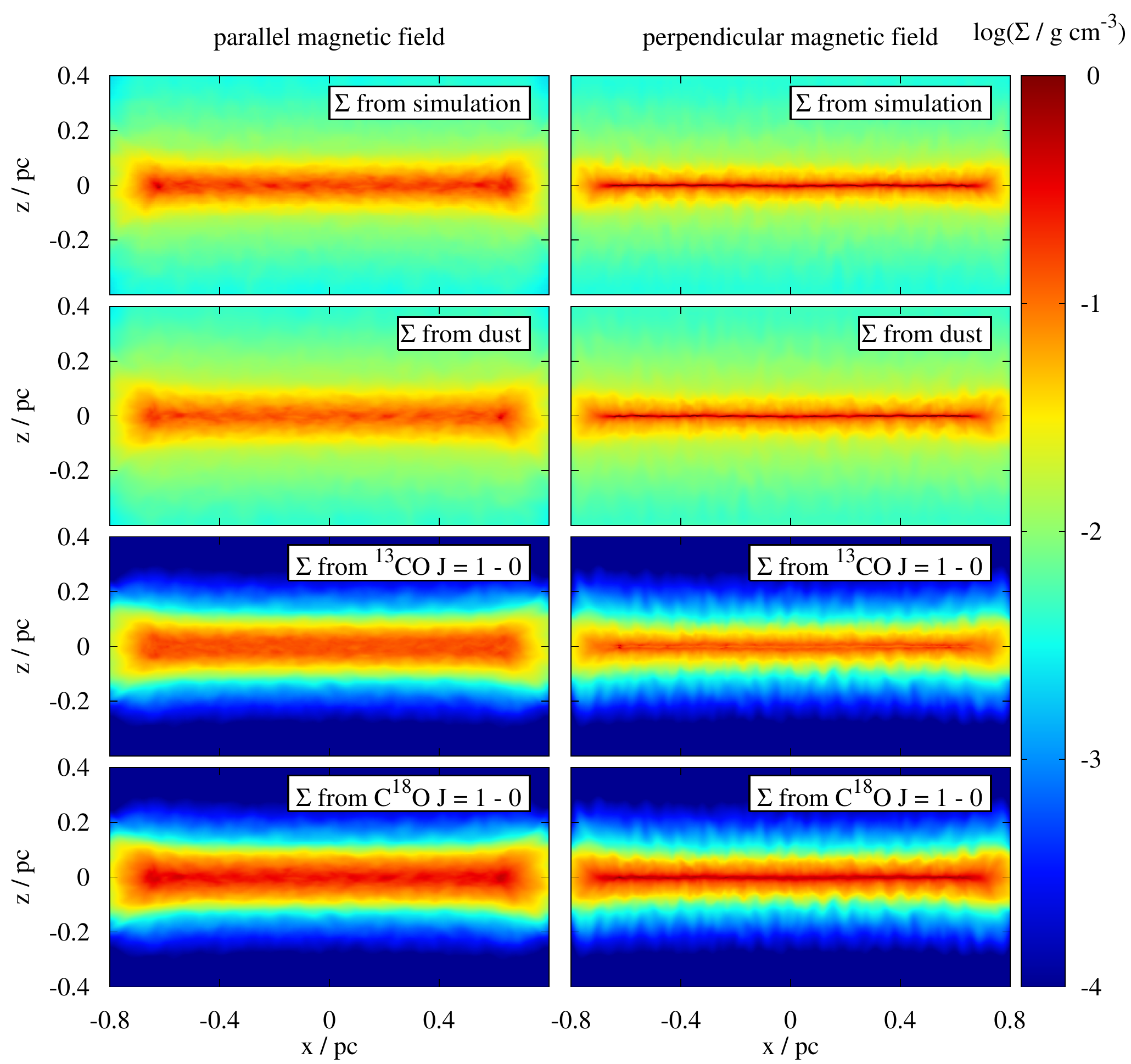}
 \caption{Column density derived from the actual simulation data (top row), dust emission (second row), the $^{13}$CO, $J$ = 1 -- 0 line (third row), and the C$^{18}$O, $J$ = 1 -- 0 line (bottom row) for the runs with a CRIR of 1 $\times$ 10$^{-16}$ s$^{-1}$ at $t$ = 300 kyr. Both a parallel (left column) and a perpendicular magnetic field configuration (right column) are considered.}
 \label{fig:coldens}
\end{figure}

In Fig.~\ref{fig:coldens} we exemplarily show the total column densities derived from the simulation data, dust emission, as well as the $^{13}$CO, $J$ = 1 -- 0 and the C$^{18}$O, $J$ = 1 -- 0 line, for the run with a CRIR of 1 $\times$ 10$^{-16}$ s$^{-1}$ at $t$ = 300 kyr. The column density derived from the dust emission (second row) is remarkably similar to the actual one (top panel). Furthermore, the column densities derived from  both $^{13}$CO and C$^{18}$O differ significantly in the outer parts of the filament, which we attribute to the lower fractional abundances of CO in these regions. However, in particular in the inner parts, which dominate the determination of the mass and width, the column densities look similar. We will discuss these results more quantitatively in the following.

In order to avoid potential confusion with various ways to fit an analytical profile to the observed profile (e.g. a Plummer or Gaussian profile), we here simply use the full-width-half-maximum (FWHM) from the radial column density profiles as a measure of the filament width. We note that in order to smooth out local fluctuations, we average the profiles along the filament axis in the inner 1.2 pc. This is somewhat smaller than the initial length of the filaments of 1.6 pc in order to avoid effects from the overdense knots appearing at the edges of the filaments \citep{Pon11,Pon12,Seifried15}. The masses, on the other hand, are inferred from the entire map ranging over 1.6 pc $\times$ 0.8 pc.

We then compare the \textit{observed} values for the total gas mass and the filament width to the \textit{actual} ones directly inferred from the simulations. The actual mass is determined from the actual column density obtained from the simulations in each pixel by summing over the same pixels as done in the synthetic column density maps. The actual FWHM is determined from the radial column density profile averaged along the filament axis in the inner 1.2 pc, just like for the observed FWHM.

We consider simulations with a CRIR of 1.3 $\times$ 10$^{-17}$, \mbox{1 $\times$ 10$^{-16}$}, and 1 $\times$ 10$^{-15}$ s$^{-1}$, a G$_0$ of 1.7, and both a parallel and perpendicular magnetic field configuration (first 6 runs listed in Table~\ref{tab:models}). We emphasise that in this section and Section~\ref{sec:LOS} we assume that only observations of either $^{13}$CO or C$^{18}$O are available and we thus correct for $\tau$ via Eqs.~\ref{eq:NCO} and~\ref{eq:tau} (see Section~\ref{sec:opacmulti} for the simultaneous observations of both isotopes). We focus on the initial and the final stage, i.e. $t$ = 0 and \mbox{300 kyr} for two reasons. First, at $t$ = 0, the physical structure, i.e. the density as well as the velocity, is \textit{identical for all filaments}, i.e. all considered values of the CRIR. However, since prior to that we have evolved the chemistry (but not the hydrodynamics) for 500 kyr, the chemical composition along with $T_\rmn{gas}$ and $T_\rmn{dust}$ differs. Hence, we can directly infer the impact of the CRIR on the chemistry -- which in turn influences the observationally inferred parameters -- without the need to consider a potential dynamical influence. In addition, we do not have to differentiate between the parallel and perpendicular magnetic field cases at $t$ = 0. Doing the same analysis for $t$ = 300 kyr then allows us to also investigate the dynamical impact of the CRIR and the magnetic field and how this is reflected in observations.

\subsection{Choice of the excitation temperature}
\label{sec:extemp}

As demonstrated in Fig.~\ref{fig:temp}, due to the influence of the CRIR and the different magnetic field configurations, the typical gas temperatures in the filaments vary between $\sim$ 10 and \mbox{35 K} \citep[but see also Fig. 3 in][]{Seifried16}. A higher CRIR results in higher values of $T_\rmn{gas}$, which is due to CR induced dissociation reactions, which significantly heat the gas, in particular in the inner part of the filament. Moreover, the gas experiences additional heating by a strong accretion shock around 0.01 - 0.1 pc.

\begin{figure*}
 \includegraphics[width=\linewidth]{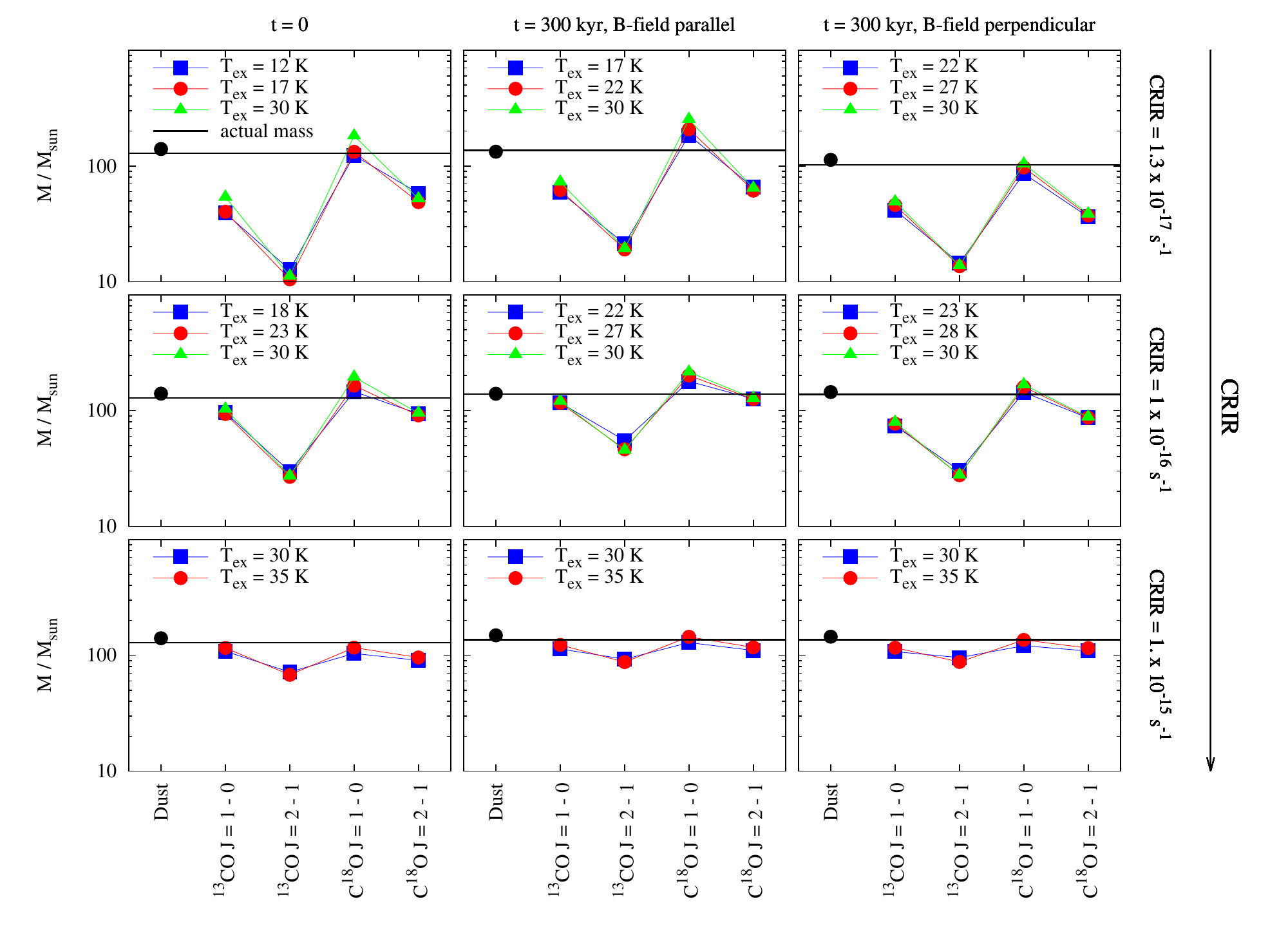}
 \caption{Filament masses inferred via the continuum emission observations (black dots) and via the four line transitions for two different values of $T_\rmn{ex}$ (blue and red dots, see Section~\ref{sec:extemp}). The values for $T_\rmn{ex}$ = 30 K in the top and middle row (green lines) are included in order to directly compare all panels. We draw lines between the red, blue, and green dots to guide the readers eye. The black lines show the actual mass. The CRIR increases from top to bottom ranging from 1.3 $\times$ 10$^{-17}$ over 1 $\times$ 10$^{-16}$ to 1 $\times$ 10$^{-15}$ s$^{-1}$. The left column shows the situation at the beginning of the simulations, and the middle and right columns at $t$ = 300 kyr for the parallel and perpendicular magnetic field case, respectively. Note that the physical structure at $t$ = 0 (left column), i.e. the actual mass (black lines), is identical for all three CRIRs.}
 \label{fig:CRIR_mass}
\end{figure*}

In actual observations of a single line, the value of $T_\rmn{ex}$ is not known, and thus a certain $T_\rmn{ex}$ is assumed to determine the CO column density \citep[e.g.][]{Hernandez11,Arzoumanian13,Hacar13}. We emphasise that, when progressively decreasing the value for $T_\rmn{ex}$, at some point, for a fixed $T_\rmn{B}$, Eq.~\ref{eq:tau} does not yield any solution for $\tau$ any more. In order account for this and still follow the observers approach, here we use the lowest possible value for $T_\rmn{ex}$, where Eq.~\ref{eq:tau} is solvable in any pixel of the emission map. Hence, $T_\rmn{ex}$ is different for different snapshots. For example, for the run with a CRIR of \mbox{1.3 $\times$ 10$^{-17}$ s$^{-1}$} and a parallel magnetic field at $t$ = 300 kyr, the bulge of material close to the centre has a temperature of $\sim$ 15 K, even though a substantial amount of gas has temperatures close to 10 K (see Fig.~\ref{fig:temp}). Since for $T_\rmn{ex}$ = 15 K still a significant number of pixels do not allow to solve Eq.~\ref{eq:tau} for $\tau$, we had to slightly increase the value of the assumed $T_\rmn{ex}$ to 17 K.

We also test the dependence of our results on the chosen value of $T_\rmn{ex}$ by using a value which is 5 K higher than the default one \citep[see e.g.][for an observational analogue]{Hernandez11}. Furthermore, in order to be able to directly compare the different snapshots, we analyse all snapshots assuming $T_\rmn{ex}$ = 30 K. The values of $T_\rmn{ex}$ used for each snapshot are shown in the corresponding figures. We also note that using a constant value for $T_\rmn{ex}$ can have severe effects on the mass estimate, which we discuss in more detail in Section~\ref{sec:temp}. A more accurate value could be obtained by observing the same line for two isotopes. Possible complications with this approach are discussed in Section~\ref{sec:opacmulti}.

Finally, we emphasise that the value of $T_\rmn{ex}$ could in principle be determined via the $^{12}$CO line emission maps, using Eq.~\ref{eq:tau} under the assumption of a very high $\tau$. However, we find that this method provides somewhat too low values of $T_\rmn{ex}$, such that Eq.~\ref{eq:tau} does not yield any solution for $^{13}$CO and C$^{18}$O any more. This is due to the fact that the optical thick $^{12}$CO emission rather probes the foreground, where $T_\rmn{ex}$ is lower than in the regions further inside, which are probed by $^{13}$CO and C$^{18}$O.

\subsection{Filament masses}
\label{sec:masses}

We first consider the inferred masses shown in Fig.~\ref{fig:CRIR_mass}, which were obtained by summing up the masses in all pixels $m_\rmn{pixel}$ (Eq.~\ref{eq:mass}). We plot the masses for the continuum emission (black dots) and for the four different line transitions (red and blue dots) using the two values of $T_\rmn{ex}$ explained in Section~\ref{sec:extemp}. To guide the readers eye, we connect the points by lines and plot the actual masses obtained from the simulation data with black lines. The left column shows the results for $t$ = 0, where the parallel and perpendicular magnetic field case yield the same results since no dynamical evolution had taken place yet. The middle and right columns show the results at $t$ = 300 kyr for the parallel and perpendicular magnetic field cases, respectively.

The mass estimates obtained from the continuum emission (black dots in Fig.~\ref{fig:CRIR_mass}) match the actual masses very accurately within a few percent \citep[see also][]{Juvela12b}, although they tend to slightly overestimate the actual mass. They seem to be almost unaffected by the CRIR (first column of Fig.~\ref{fig:CRIR_mass}), which can be attributed to the fact that CRs do not have a direct impact on the dust temperature. Overall, our results thus indicate a high fidelity of filament masses obtained via dust emission observations \citep[e.g.][]{Andre10,Arzoumanian11,Palmeirim13,Juvela12a,Kainulainen13,Kainulainen16b}

The masses obtained from the line emission maps show only a moderate dependence on $T_\rmn{ex}$ with variations of a few 10\% at most (for an individual line and snapshot considered), even when $T_\rmn{ex}$ is varied by a factor of $\sim$ 2 (top and middle rows of Fig.~\ref{fig:CRIR_mass}). For a given snapshot there are, however, significant variations of up to a factor of $\sim$ 10 between the masses obtained from the different lines. In particular, the mass estimates for the $J$ = 2 -- 1 transition appear much less accurate than that for the $J$ = 1 -- 0 transition, underestimating the actual mass by up to a factor of $\sim$ 10. This is in good agreement with recent observations of \citet{Nishimura15}, who find a similar trend with mass differences of a factor of $\sim$ 3. We investigate this problem in more detail in Sections~\ref{sec:temp} and~\ref{sec:subtherm}. Furthermore, the masses inferred from the $^{13}$CO lines tend to underestimate the actual masses and are also lower than the corresponding C$^{18}$O masses, which are closer to the actual value. We attribute this to the higher opacities for $^{13}$CO, which seemingly cannot fully be accounted for via Eq.~\ref{eq:tau}, possibly due to the highly non-linear effects of radiative transfer. This underestimation for $^{13}$CO is in good agreement with previously reported results of \citet{Padoan00} and \citet{Szucs16}.

\begin{figure}
 \includegraphics[width=\linewidth]{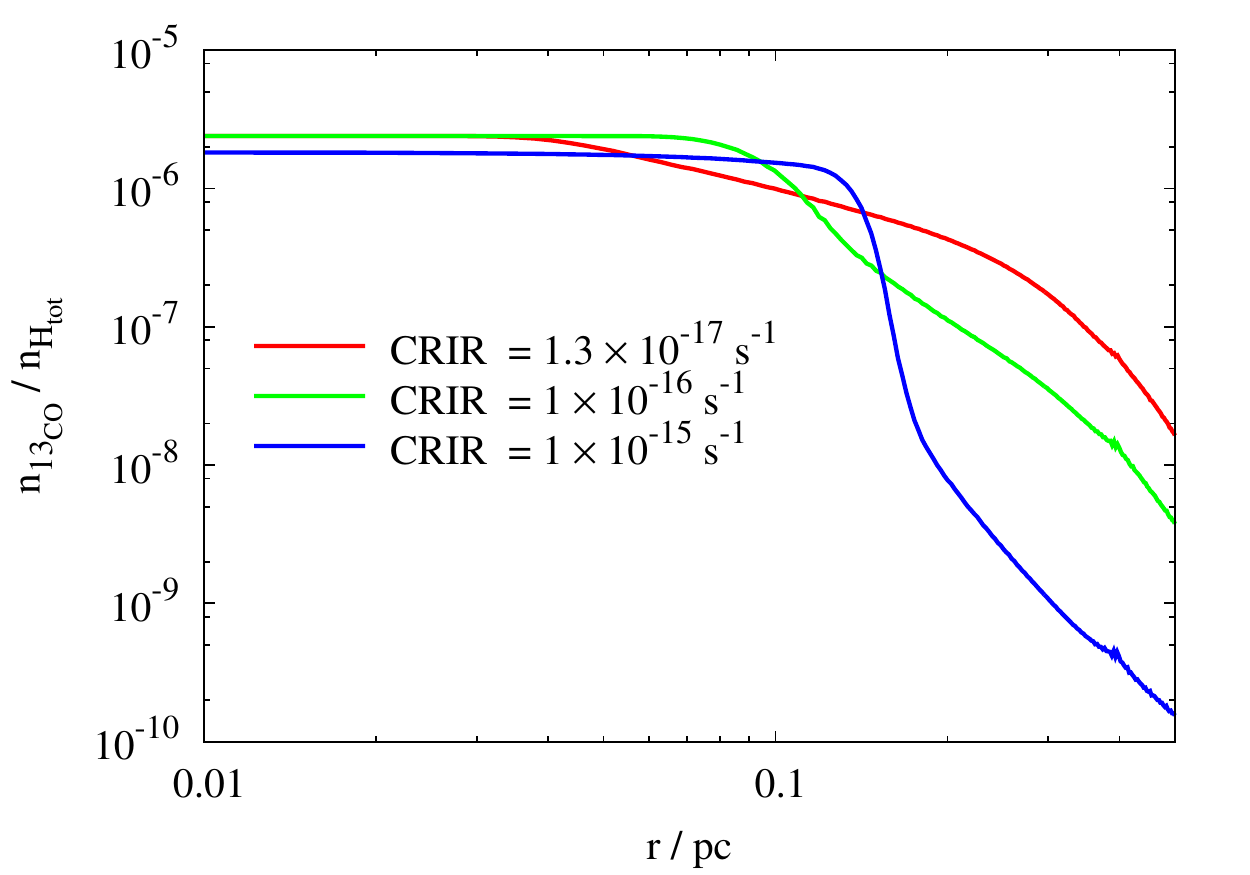}
 \caption{Radial profile of the number density of $^{13}$CO normalised to the total hydrogen number density at $t$ = 300 kyr for the filaments with a parallel magnetic field and three different CRIRs. The $^{13}$CO number density in the outer parts decreases with increasing CRIR due to the enhanced dissociation of CO by CRs.}
 \label{fig:COprofile}
\end{figure}

Next, we consider the impact of the CRIR. Overall, as the CRIR increases (from top to bottom in Fig.~\ref{fig:CRIR_mass}), the mass estimates get more accurate and the relative differences between the four different lines transitions for a given snapshot decrease. Interestingly, for $^{13}$CO the inferred masses seem to increase with an increasing CRIR, although the actual mass of CO in the filament (obtained from the simulation data directly) is decreasing. We attribute this effect to the fact that due to the higher destruction rate of CO for higher CRIRs, $^{13}$CO is less abundant in the outer part of the filament (see Fig.~\ref{fig:COprofile}). This reduces the self-absorption along the LOS and increases the inferred mass (see Section~\ref{sec:subtherm}).

Finally, comparing the middle and right column of Fig.~\ref{fig:CRIR_mass}, we find that the magnetic field configuration has very little influence on the actual mass. We attribute this to the fact that we cover the entire extent of the filament (see Fig.~\ref{fig:coldens}), and thus the total mass contained in the considered regions is approximately constant\footnote{We note that the somewhat smaller actual mass in the top-right panel of Fig.~\ref{fig:CRIR_mass} is due to the fact that in this run we made use of sink particles \citep{Federrath10}, i.e. some amount of the gas was removed from the simulation grid and stored in these sink particles which only interact gravitationally with the gas. This was necessary since -- due to the rapid collapse of the filament -- the gas very quickly reaches very high densities. In case no sink particles were used, this would prevent us from simulating the evolution over a few 100 kyr.}. In contrast to that, the radial profiles are strongly affected by the field configuration \citep{Seifried15,Seifried16}, which in turn could affect the emission properties and thus the \textit{derived} masses. However, due to the large differences between the four different CO lines (for an individual snapshot), it is difficult to infer an impact of the radial profile (and thus the field configuration) on the observed line emission masses. The dust masses are even less affected, since the continuum emission remains optically thin.

\subsection{Filament widths}
\label{sec:width}

\subsubsection{Line emission}
\label{sec:width_line}

\begin{figure*}
 \includegraphics[width=0.8\linewidth]{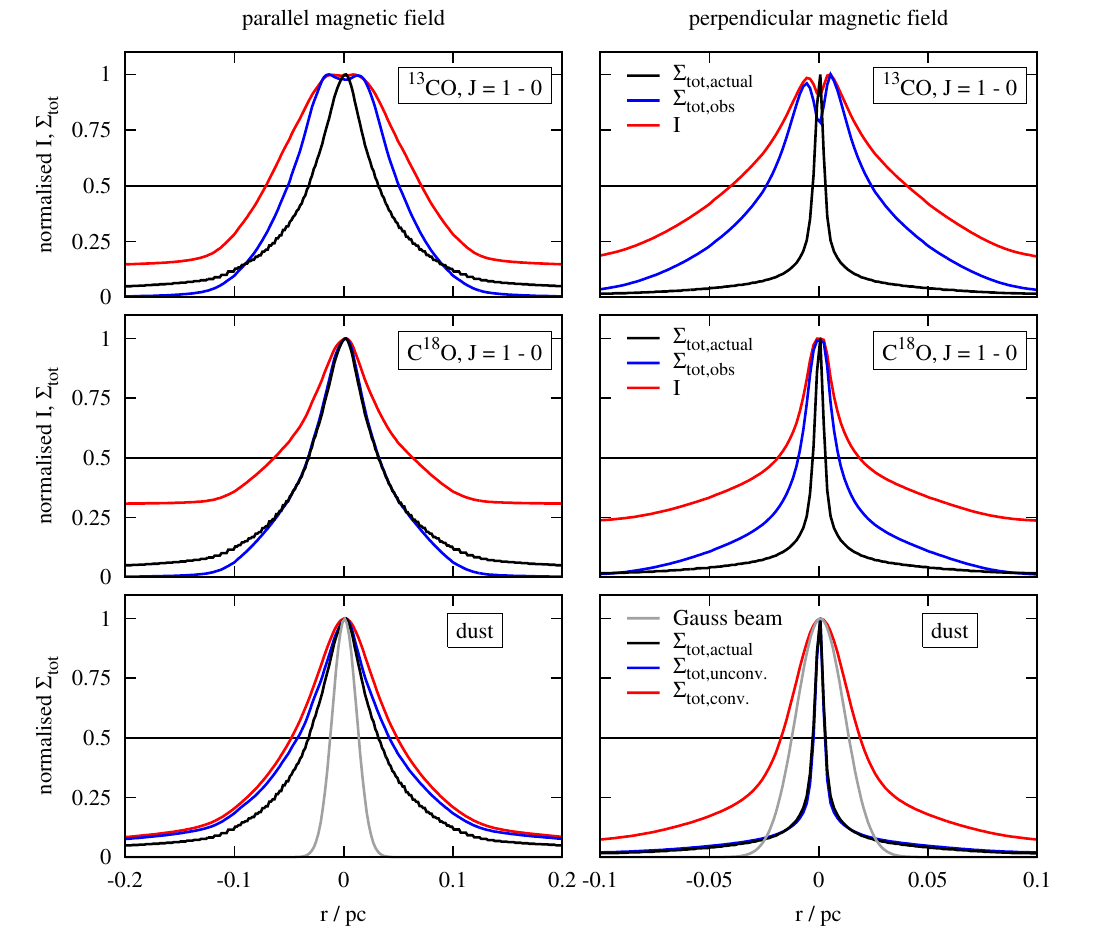}
 \caption{Radial profiles for the runs with a CRIR of \mbox{1 $\times$ 10$^{-16}$ s$^{-1}$} at $t$ = 300 kyr for the parallel (left column) and perpendicular magnetic field case (right column). All profiles are normalised with respect to their maximum value. Each intersection point with the black horizontal line indicates the corresponding FWHM. Top and middle row:  Integrated intensity of the CO emission (red lines), the total gas column density derived from it (blue lines), and the actual gas column density obtained from the simulation data (black lines). Bottom row: Total column density obtained from the dust emission for the case without (blue lines) and with  convolution with a Gaussian beam (red lines). The Gaussian beam used is shown with the grey line (see text). Please note the different radial range in the left and right column.}
\label{fig:profiles}
\end{figure*}

Next, we determine the FWHMs of the filaments for the same runs as in the previous section. In the two upper rows of Fig.~\ref{fig:profiles}, we plot the normalised radial profile of the integrated intensity of the $J$ = 1 -- 0 transition of both $^{13}$CO and C$^{18}$O and the inferred total gas column density for the runs with a CRIR of 1 $\times$ 10$^{-16}$ s$^{-1}$ at $t$ = 300 kyr, i.e. for the same maps as shown in Fig.~\ref{fig:coldens}. The column density is inferred using \mbox{$T_\rmn{ex}$ = 22 (23) K} for the parallel (perpendicular) magnetic field case. We note that for the corresponding $J$ = 2 -- 1 transitions qualitatively very similar results are obtained.
Some interesting results can be read off. First, the profiles of the integrated intensity are wider than that of the corresponding column density maps. This is due to the fact that the opacity correction increases the central column density, thus decreasing the apparent width. Secondly, the profiles obtained for $^{13}$CO are wider than that for C$^{18}$O, which we attribute to the higher optical depth of the $^{13}$CO transitions. Finally, both the intensity and the column density profiles for $^{13}$CO show a dip towards the centre of the filament, which we attribute to various LOS effects (see Section~\ref{sec:LOS}.)

\begin{figure*}
 \includegraphics[width=\linewidth]{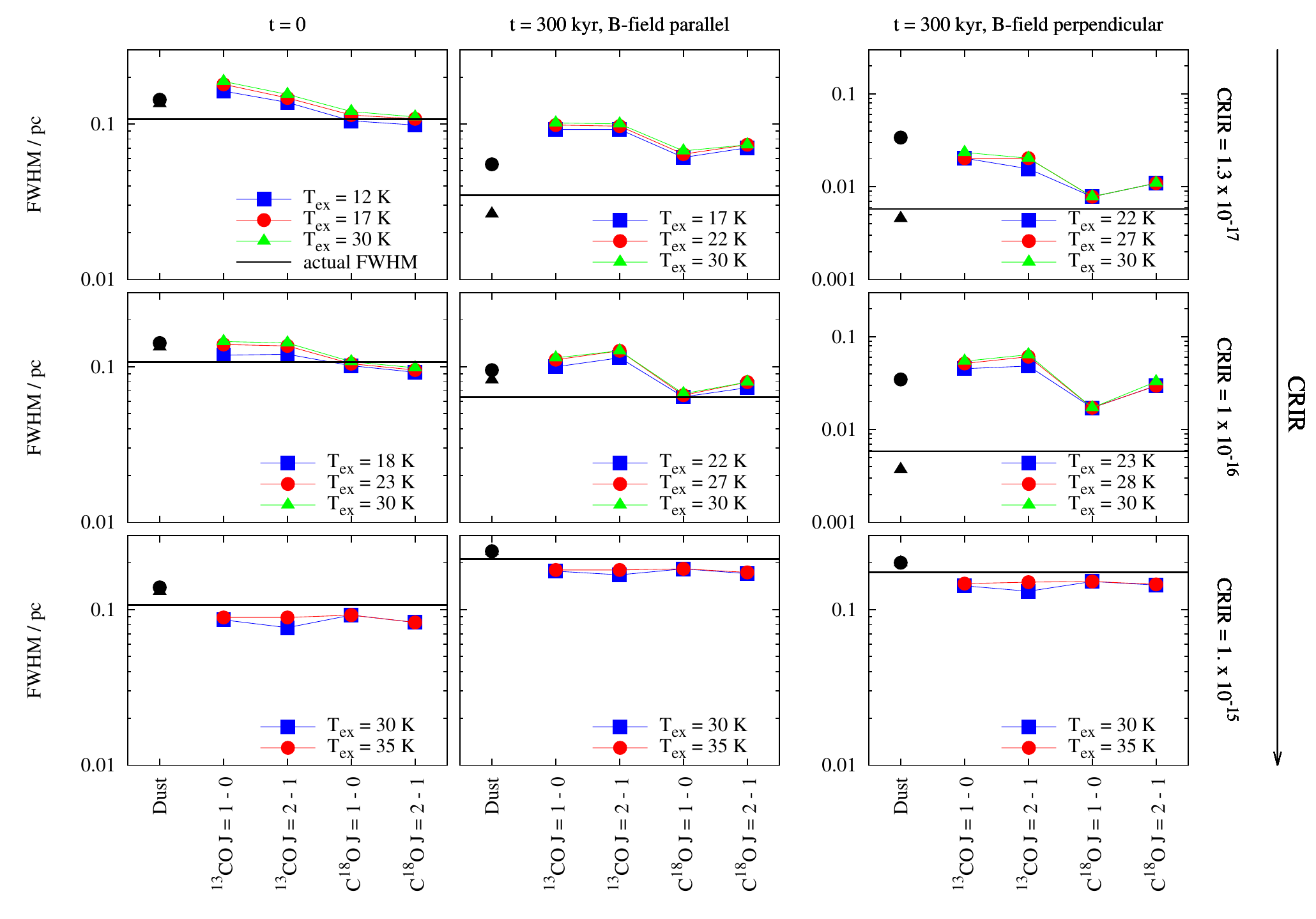}
 \caption{Inferred FWHMs of the filaments. The order of the snapshots is identical to Fig.~\ref{fig:CRIR_mass}. Black circles and triangles show the FWHMs obtained from continuum emission observations smoothed with a 36'' beam as e.g. for \textit{Herschel} data (assuming a distance of 150 pc) and without any resolution limitations, respectively. Red, blue, and green dots show the FWHMs obtained from line emission maps, and black lines the actual ones. The inferred FWHMs for condensed filament are partly significantly overestimated.}
\label{fig:CRIR_width}
\end{figure*}

Next, in Fig.~\ref{fig:CRIR_width}, we plot the FWHMs inferred from the column density profiles for both the line and continuum emission for all snapshots considered. The order of the snapshots is identical to that in Fig.~\ref{fig:CRIR_mass}. The black line shows the actual FWHM, the black dots that obtained from the dust emission maps, and the red, blue, and green dots those from the line emission maps for the various values of $T_\rmn{ex}$ used (Section~\ref{sec:extemp}) .

Similar to the masses, the FWHMs obtained via line emission observations depend only weakly on the chosen $T_\rmn{ex}$ (variations by a few 10\%). However, we do not find the systematic differences between the $J$ = 1 -- 0 and $J$ = 2 -- 1 transitions. The results also confirm the findings in Fig.~\ref{fig:profiles} that $^{13}$CO gives somewhat larger values for the FWHM than C$^{18}$O. Furthermore, for an actual FWHM around 0.1 pc (compare Table~\ref{tab:models}), the inferred FWHMs match reasonably well within a factor of 2 or less, although -- except for the highest CRIR -- the values for $^{13}$CO tend to slightly overestimate the actual FWHM (black line). 

Considering the upper middle, upper right, and middle right panels of Fig.~\ref{fig:CRIR_width}, we find that these filaments have rather small FWHMs ($<$ 0.05 pc; see also Table~\ref{tab:models}). For this reason, throughout the paper we will refer to them as \textit{condensed filaments}. Most interestingly, for these condensed filaments the FWHMs inferred from both CO isotopes are in general well above the actual FWHMs. This indicates a potential problem of line emission observations to accurately measure very narrow filament widths.

As can be read off from Table~\ref{tab:models}, small FWHMs seem to preferentially occur for low CRIRs and/or a perpendicular magnetic field configuration.  This can be understood as a consequence of a reduced pressure support against radial collapse of the filament: For low CRIRs the thermal pressure is lower due to the reduced amount of CR heating (see Fig.~\ref{fig:temp}), and for perpendicular magnetic fields no radial magnetic pressure gradient is present \citep[see also][]{Seifried15,Seifried16}. Furthermore, since the profiles for the integrated intensities are even wider (Fig.~\ref{fig:profiles}) and thus possibly even more inaccurate, we strongly suggest to infer filament widths from the opacity corrected column density maps instead of the intensity maps \citep[e.g.][]{Panopoulou14}.

Finally, considering the effect of the CRIR (\mbox{$t$ = 0,} left column of Fig.~\ref{fig:CRIR_width}), we find that the inferred FWHMs decrease with an increasing CRIR, and for CRIR = \mbox{1 $\times$ 10$^{-15}$ s$^{-1}$} the FWHMs are systematically underestimated. We attribute this to the fact that for a high CRIR, CO is dissociated in the outer parts of the filament (see Fig.~\ref{fig:COprofile}) leading to a (apparent) smaller width.

We note that instead of directly determining the FWHM from the profiles, we also tried to fit a Plummer or Gaussian function to the profile and subsequently determine the FWHM from the fit. However, this does not result in better matches, even when excluding the inner part of the profile from the fit, where self-absorption is taking place. This is due to the wings of the profiles, which do not have a Gaussian or Plummer-like shape. In fact, for the condensed filaments the fit rather results in FWHMs which are even further off from the actual ones.

\subsubsection{Dust emission}
\label{sec:width_dust}

Next, we consider the results obtained from the dust emission maps. Assuming a distance of 150 pc typical for nearby low-mass star-forming regions (e.g. the Taurus molecular cloud), the resolution of 0.0016 pc of the emission maps corresponds to about 2''. This, however, is significantly higher than the resolution of e.g. the \textit{Herschel} or IRAM telescopes working in the infrared and millimetre regime. Hence, in order to make predictions for actual observations with these instruments, we convolve the emission maps with a Gaussian beam with a size of 36'' corresponding to the beam size of \textit{Herschel} at 500 $\mu$m \citep[see][for the effect of noise]{Juvela12b}. We use all simulated wavelengths from 70 $\mu$m to 2.6 mm and first consider the convolved column density profiles for two filaments in the bottom row of Fig.~\ref{fig:profiles} (red lines). No dips towards the centre can be seen, which is due to the fact that the dust remains optically thin. However, for the perpendicular magnetic field case (bottom right panel), the obtained profile is dictated by the Gaussian beam (grey line) and thus significantly off from the actual one (black line).

The inferred FWHMs for all snapshots are shown in Fig.~\ref{fig:CRIR_width} with black circles. Except for the three very condensed filaments, the actual FWHMs are reasonably well reproduced with differences of at most 30\%. For the three condensed filaments, however, the obtained FWHMs are dictated by the imposed resolution limit of 0.026 pc and overestimate the actual FWHMs by a factor of a few \citep[see also][]{Juvela12b}, similar to the results obtained from the line emission maps. Hence, it seems that neither line nor continuum emission observations (with current single dish telescopes) can accurately probe the width of these condensed filaments but rather give apparent widths of a few times 0.01 pc to 0.1 pc. We emphasise that this lower limit is in good agreement with recent observations \citep{Juvela12a,Malinen12,Sanchez14}.

However, when using spatial filtering as suggested by \citet{Palmeirim13} or the Atacama Large Millimeter/submillimeter Array, also higher spatial resolutions of 1'' or below are now accessible in the infrared and millimetre regime. For this reason, in the bottom row of Fig.~\ref{fig:profiles}, for the selected snapshots we also show the column density profiles obtained from the dust emission maps without convolution. In particular for the condensed filament (perpendicular magnetic field, bottom right panel), the match of the observed profile is now significantly improved. This is also reflected in Fig.~\ref{fig:CRIR_width}, where the obtained FWHMs for all runs, under the assumption of no observational resolution limitations, are shown with black triangles: For the filaments with a FWHM around 0.1 pc, there are only marginal differences of $\leq$ 10\% compared to the convolved case. For the condensed filaments, however, the match between the actual and observed FWHMs seems to be somewhat improved, although there are indications that the observed value is now slightly too low. We note, however, that here we did not consider any interferometric effects as we have done in \citet{Seifried16b}, which is why the accuracy of these predictions for actual observations is limited.

\section{Line-of-sight effects}
\label{sec:LOS}

\subsection{Temperature variations}
\label{sec:temp}

As shown in Fig.~\ref{fig:temp} \citep[but see also][]{Seifried16}, there are significant variations of $T_\rmn{gas}$ along the radial direction, which is a consequence of the heating by CRs as well as the accretion shock occurring at a radius of 0.01 -- 0.1 pc. Hence, the assumption of a fixed $T_\rmn{ex}$ along the LOS, which is often made in actual observations \citep[e.g.][]{Hernandez11,Arzoumanian13,Hacar13}, is most likely an oversimplification, which can contribute to some of the deviations between the derived and actual masses and FWHMs discussed before (compare Figs.~\ref{fig:CRIR_mass} and~\ref{fig:CRIR_width}).

\begin{figure}
 \includegraphics[width=\linewidth]{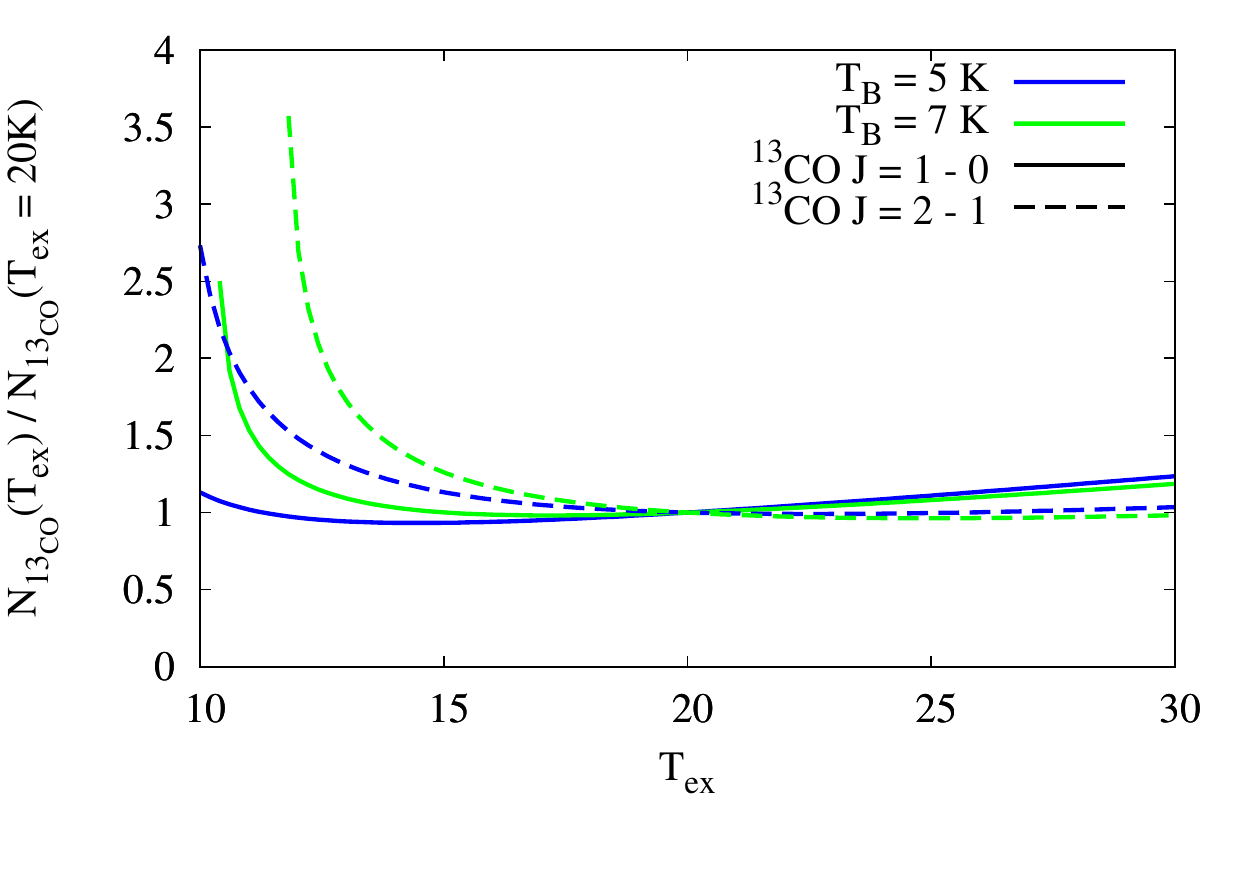}
 \caption{Effect of the assumed excitation temperature $T_\rmn{ex}$ on the derived $^{13}$CO column density (Eq.~\ref{eq:NCO}) assuming typical brightness temperatures of 5 K (red lines) and 7 K (green lines) found in our synthetic emission maps. We normalise the values to the value of $N_\rmn{^{13}CO}$ at $T_\rmn{ex}$ = 20 K. Towards lower $T_\rmn{ex}$, $N_\rmn{^{13}CO}$ increases faster for the $J$ = 2 -- 1 transition (dashed lines) than for the $J$ = 1 -- 0 transition.}
 \label{fig:Tex}
\end{figure}

We investigate this more quantitatively in Fig.~\ref{fig:Tex}, where we plot the dependence of the (normalised) $^{13}$CO column density (Eq.~\ref{eq:NCO}) on the assumed excitation temperature using typical brightness temperatures found in our emission maps\footnote{Note that for C$^{18}$O, which is not shown here, the results are quantitatively and qualitatively very similar.}.
For the $J$ = 2 -- 1 transition (dashed lines), $N_\rmn{^{13}CO}$ shows a significantly stronger dependence on $T_\rmn{ex}$ than for the $J$ = 1 -- 0 transition (solid lines): Reducing $T_\rmn{ex}$ from a fiducial value of 20 K to 12 K would increase $N_\rmn{^{13}CO}$ by a factor of $\sim$ 1.5 - 3.5 for $J$ = 2 -- 1, whereas for $J$ = 1 -- 0 the increase is only marginal (a few 10\%). This is a consequence of the occurrence of the transition frequency $\nu$ and the upper state energy $E_\rmn{u}$ in the \textit{exponent} in Eqs.~\ref{eq:NCO} and~\ref{eq:tau}. Given the fact that $T_\rmn{gas}$ varies significantly along the LOS (down to about 10 K; see Fig.~\ref{fig:temp}), some of the considered velocity channels in our synthetic observations most likely have a lower $T_\rmn{ex}$ than the actually assumed one.

Hence, the masses obtained from the $J$ = 2 -- 1 transition can be easily lower than the actual \textit{and} the $J$ = 1 -- 0 mass by a factor of $\sim$ 2 or more, which -- at least in parts -- explains the peculiar behaviour found in Fig.~\ref{fig:CRIR_mass} for the two lower CRIRs. Furthermore, for the runs with a CRIR of \mbox{1 $\times$ 10$^{-15}$ s$^{-1}$} the differences between the masses obtained from the $J$ = 1 -- 0 and $J$ = 2 -- 1 lines are significantly smaller (bottom row). This fits nicely with the fact that for the excitation temperatures applied here ($>$ 20 K), $N_\rmn{^{13}CO}$ is much less sensitive to $T_\rmn{ex}$ than at low $T_\rmn{ex}$ (Fig.~\ref{fig:Tex}) for both transitions. This also explains the small differences in the masses for different $T_\rmn{ex}$ for a \textit{single} transition since only excitation temperatures above 15 K were assumed (see Fig.~\ref{fig:CRIR_mass}).

The assumption of a constant $T_\rmn{ex}$ possibly also contributes to the overestimation of the FWHMs inferred from the CO emission maps for the condensed filaments (compare Fig.~\ref{fig:CRIR_width}). Considering the radial profiles of the total column density $\Sigma_\rmn{tot,obs}$ (blue lines in the right column of Fig.~\ref{fig:profiles}), they appear significantly wider than the actual column density. Since, however, $T_\rmn{gas}$ drops towards the centre of the filament (compare Fig.~\ref{fig:Tex}), the effect described in Fig.~\ref{fig:temp} results in an underestimation of the central column density, which seems to be particularly pronounced for the condensed filaments. This in turn results in apparently wider profiles, thus contributing to the overestimation of the FWHMs.

\begin{figure}
 \includegraphics[width=\linewidth]{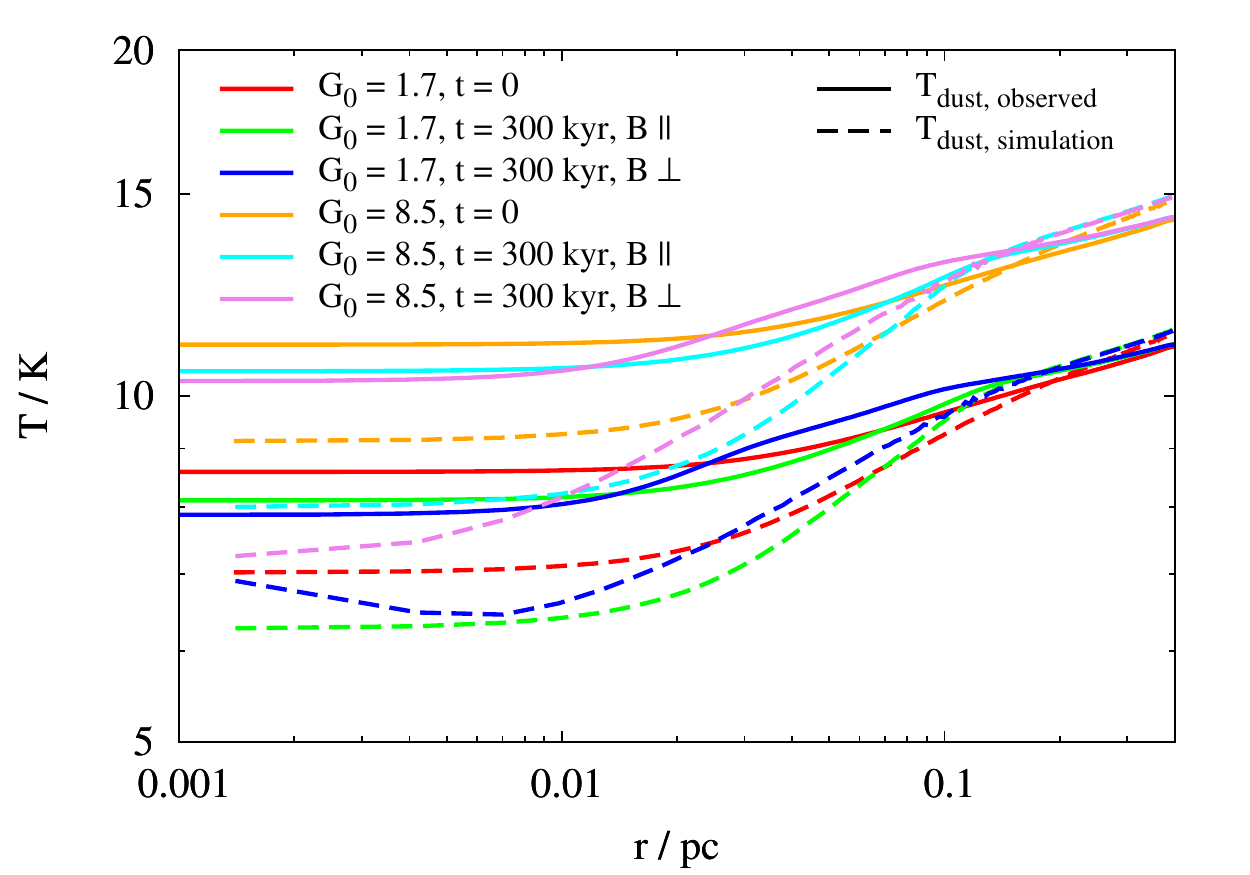}
 \caption{Observed (solid lines) and actual (dashed lines) dust temperature in the simulations with a CRIR of 1 $\times$ 10$^{-16}$ s$^{-1}$ and two different values of $G_0$. Both the observed and actual profiles follow a rough polytropic relation. The observed $T_\rmn{dust}$ is somewhat higher than the actual one due to warmer dust in the foreground.}
 \label{fig:dusttemp} 
\end{figure}

In this context we again note that the \textit{dust} temperature roughly follows a polytropic relation $T_\rmn{dust} \propto \rho^{\gamma-1}$ with \mbox{$\gamma$ =  0.9 - 0.95} \citep[][]{Seifried16} in agreement with recent \textit{Herschel} observations \citep{Arzoumanian11,Palmeirim13,Li14}. In Fig.~\ref{fig:dusttemp} we investigate how accurately one can measure $T_\rmn{dust}$ using the runs with a CRIR of 1 $\times$ 10$^{-16}$ s$^{-1}$ and $G_0$ = 1.7 as well as 8.5. First, the simulation data (dashed lines) show that the runs with $G_0$ = 8.5 have values of $T_\rmn{dust}$ which are a few K higher due to the enhanced radiative heating of the dust grains. This is also recovered in the observed dust temperatures. Moreover, we find that the observed $T_\rmn{dust}$ profiles reproduce the actual profiles reasonably well, however, shifted upwards by at most \mbox{2 -- 3 K}.

As reported by \citet{Shetty09}, this shift is due to the fact that along the LOS gas with higher dust temperatures, in particular in the outer parts, is included \citep[see also][]{Marsh15}. Indeed, when fitting the modified blackbody spectrum only for wavelengths $\geq$ 500 $\mu$m as suggested by \citet{Shetty09}, we find that the obtained values of $T_\rmn{dust}$ are lower by about 0.5 -- 1 K and thus in better agreement with the actual values.

\subsection{Subthermal excitation}
\label{sec:subtherm}

Beside the relative differences between the $J$ = 2 -- 1 and $J$ = 1 -- 0 masses for both $^{13}$CO and C$^{18}$O, in particular for $^{13}$CO the actual masses are underestimated partly by a factor of up to 10 \citep[see Fig.~\ref{fig:CRIR_mass}, but also][]{Padoan00,Szucs16}. The aforementioned temperature variations can at most account for a factor of 2 -- 3 for the $J$ = 2 -- 1 transition (and for a few 10\% for the $J$ = 1 -- 0 transition), which indicates that a second (LOS) effect might be at work.

\begin{figure}
 \includegraphics[width=\linewidth]{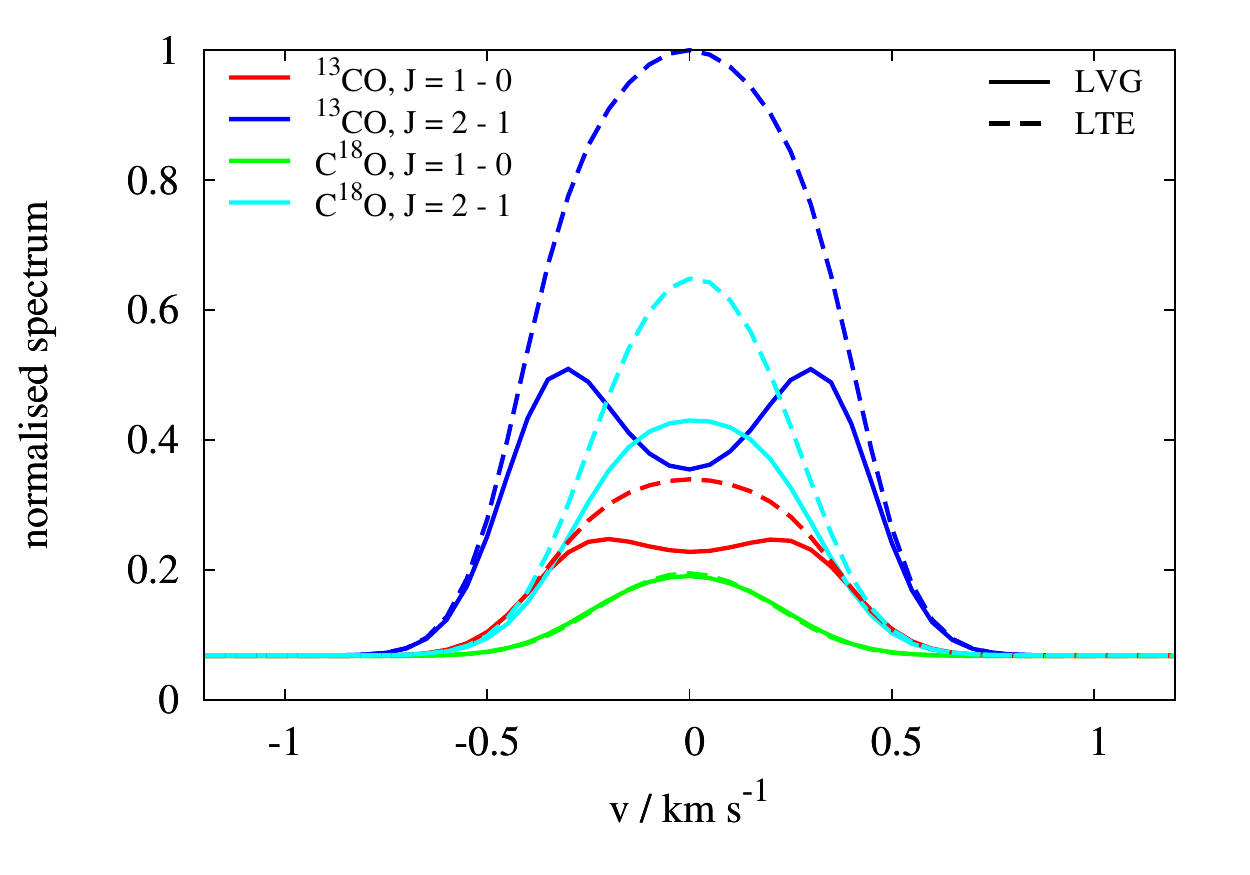}
 \caption{Normalised spectra of the $^{13}$CO and C$^{18}$O line transitions for the central part of the filament in the run with a CRIR of 1.3 $\times$ 10$^{-17}$ s$^{-1}$ at $t$ = 0. The spectra are obtained with the LVG method (used throughout the paper) and for comparison also under the assumption of LTE. The LVG spectra are lower and show a strong self-absorption feature for $^{13}$CO towards the line centre due to subthermal excitation of $^{13}$CO in the outer parts of the filament. The C$^{18}$O spectra differ significantly from that of $^{13}$CO, which is why a channel-by-channel correction of the optical depth is problematic (see Section~\ref{sec:opacmulti}).}
 \label{fig:spectra}
\end{figure}

In Fig.~\ref{fig:spectra} we consider the spectra of the four CO transitions for the run with a CRIR of 1.3 $\times$ 10$^{-17}$ s$^{-1}$ at \mbox{$t$ = 0}. We average the spectra over a region of 1.2 pc $\times$ 0.1 pc to smooth out local fluctuations. In addition to the spectra obtained via the LVG method used throughout this paper, we also show the (hypothetical) spectra obtained when running the radiative transfer calculation under the assumption that CO is in local thermal equilibrium (LTE).

\begin{figure*}
 \includegraphics[width=\linewidth]{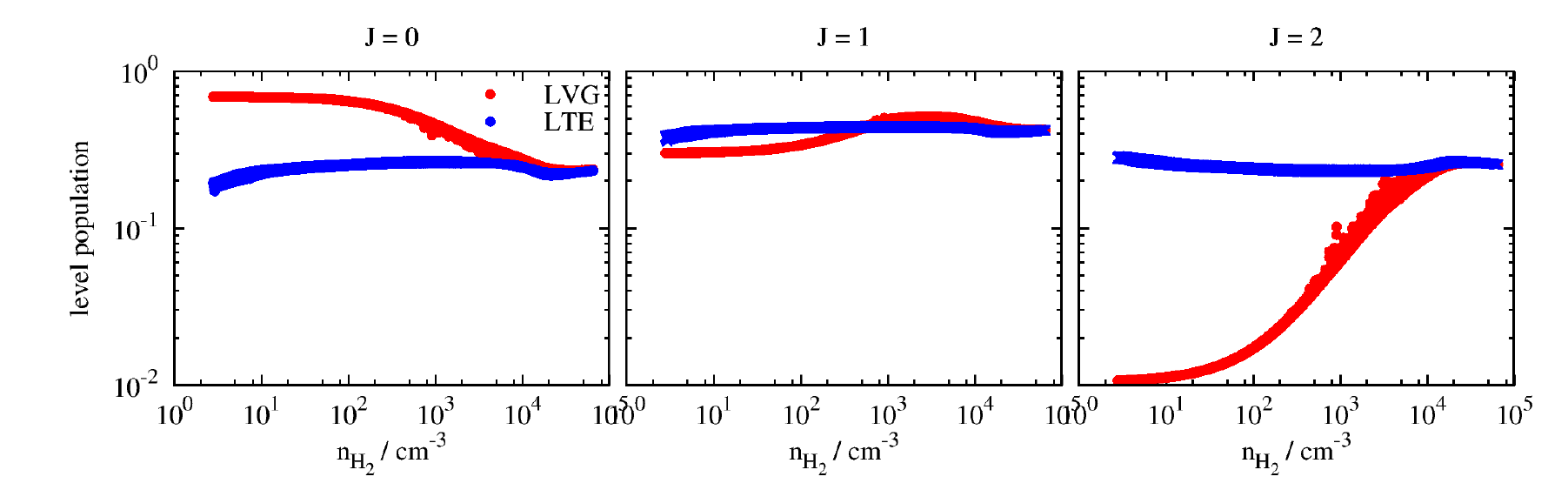}
 \caption{Dependence of the level populations of $^{13}$CO on the hydrogen number density for both the LVG (red dots) and LTE method (blue dots). For $J$ = 1 and 2, $^{13}$CO is subthermally excited in the low-density region, i.e. the outer parts of the filaments, whereas in the centre it is almost in LTE.}
 \label{fig:levelpop}
\end{figure*}

Overall, the LVG spectra are lower compared to the LTE spectra. Since in addition for $^{13}$CO they even show a strong self-absorption feature \citep[see e.g.][for corresponding observations]{Schneider10}, we attribute this to the high optical depth (see also Section~\ref{sec:opacsingle}) in the centre of the filaments. Furthermore, as suggested by \citet{Nishimura15}, who find similar differences between the $J$ = 2 -- 1 and \mbox{$J$ = 1 -- 0} masses as in this work, this could be a consequence of the subthermal excitation of CO. In order to test this, in Fig.~\ref{fig:levelpop} we compare -- for each cell of the simulation grid -- the level populations for $^{13}$CO as obtained with RADMC-3D for the LVG and the LTE method\footnote{For C$^{18}$O very similar results are obtained.}. We find that in particular in the lower density regime, i.e. the outer parts of the filament the $J$ = 1 and 2 levels are less populated than expected under the LTE assumption. This is due to the fact that the H$_2$ number density $n_\rmn{H_2}$ in this region drops below the critical density of about $2 - 3 \times 10^3$ cm$^{-3}$ for the $J$ = 1 -- 0 and about $8 - 10 \times 10^3$ cm$^{-3}$ for the $J$ = 2 -- 1 transition. Due to the higher critical density for the latter, the subthermal excitation is more pronounced for the $J$ = 2 level, which also explains the large deviation of the $J$ = 2 -- 1 masses from the true masses.

Hence, the following picture arises: in the centre of the filament CO is in LTE and emits accordingly. In the outer part (foreground), however, CO is only subthermally excited and will thus emit less than expected for LTE. Moreover, in this region any radiation (from the centre of the filament) might in parts be self-absorbed depending on the optical depth. Overall, in combination with the effect of temperature variations along the LOS (Section~\ref{sec:temp}), this effect can lower the derived masses, an effect which is particularly pronounced for the $J$ = 2 -- 1 transition.

We emphasise that this result qualitatively also holds for higher CRIRs. However, in particular the level population for $J$ =  1 increases with increasing CRIR. In addition, in the outer parts (foreground) the temperatures are higher (see Fig.~\ref{fig:temp}) and the amount of CO and thus also the strength of self-absorption are reduced. In combination, for higher CRIRs these effects result in an increasingly better match between the actual and observationally obtained masses (see Fig.~\ref{fig:CRIR_mass}).

We also note that as shown in Fig.~\ref{fig:spectra} the spectral range from -1.5 to 1.5 km s$^{-1}$ used in this work is sufficient to cover all emission.
We also carefully checked our analysis tools for potential bugs. Furthermore, for the LVG method we artificially increased the H$_2$ number density by a factor of 100. As expected, in this case the spectra match the LTE spectra reasonably well and the inferred masses agree better with the actual mass, which makes us confident that the obtained results are correct.

In summary, we suggest that large differences in the masses obtained from $J$ = 2 -- 1 and $J$ = 1 -- 0 observations could possibly be used as a diagnostic tool in observations: The differences could indicate the non-isothermal state of star-forming filaments and thus possibly the occurrence of an accretion shock in the infalling gas -- as it is the case in our simulations (Section~\ref{sec:temp}) -- but also the subthermally excited state of CO. In order to disentangle both effects, we suggest to observe typical shock tracer molecules, which would indicate the presence of an accretion shock, which, due to the chemical network used in this work, cannot be probed here.

\subsection{Opacity correction}
\label{sec:opacsingle}

So far we have automatically corrected for the optical depth by means of applying Eq.~\ref{eq:tau}. If we, however, do not correct for optical depth effects, the masses inferred from $^{13}$CO drop by a few 10\%. For C$^{18}$O the effect is somewhat less pronounced (1 -- 10\%). This is due to the fact that for the simulated filaments -- which represent typical filaments found in observations -- $^{13}$CO and in parts even C$^{18}$O becomes optically thick.

Considering the FWHMs obtained from the line emission maps, we find that the values are up to a few 10\% larger if no optical depth correction is applied, which amplifies the general trend that some of the observed FWHM overestimate the actual ones (see Fig.~\ref{fig:CRIR_width}, and Section~\ref{sec:width_line}). However, in particular for wider filaments with a FWHM around 0.1 pc (see Table~\ref{tab:models}), the opacity \textit{uncorrected} FWHMs still match the actual ones within a factor of about 2.

Checking the values of $\tau$ determined from Eq.~\ref{eq:tau} for each pixel and velocity channel, we find typical values of $\tau$ = 1 -- 5 for $^{13}$CO and $\tau$ = 0.1 -- 1 for C$^{18}$O, although for the latter maximum values of up to $\tau \sim$ 2.5 are reached locally. These derived values for $\tau$ are in good agreement with the findings of \citet[][see their Fig.~4]{Arzoumanian13}.

Overall, our results show that observations should be carefully corrected for the optical depth, although even these opacity corrected quantities can deviate from the actual values, which is particularly pronounced for the masses (see Fig.~\ref{fig:CRIR_mass}, and Section~\ref{sec:masses}).

\subsection{Inclination effects}
\label{sec:incl}

Recently, filaments with widths up to about 1 pc -- and thus significantly larger than those found with \textit{Herschel} observations \citep[e.g.][]{Arzoumanian11,Peretto12,Palmeirim13} -- were reported \citep[e.g.][]{Juvela12a,Malinen12,Panopoulou14}. Since on average filaments are expected to be inclined with respect to the plane of the sky, one possible explanation for observed differences could be that molecular line and dust emission observations have different sensitivities to inclination effects.

\begin{figure*}
 \includegraphics[width=\linewidth]{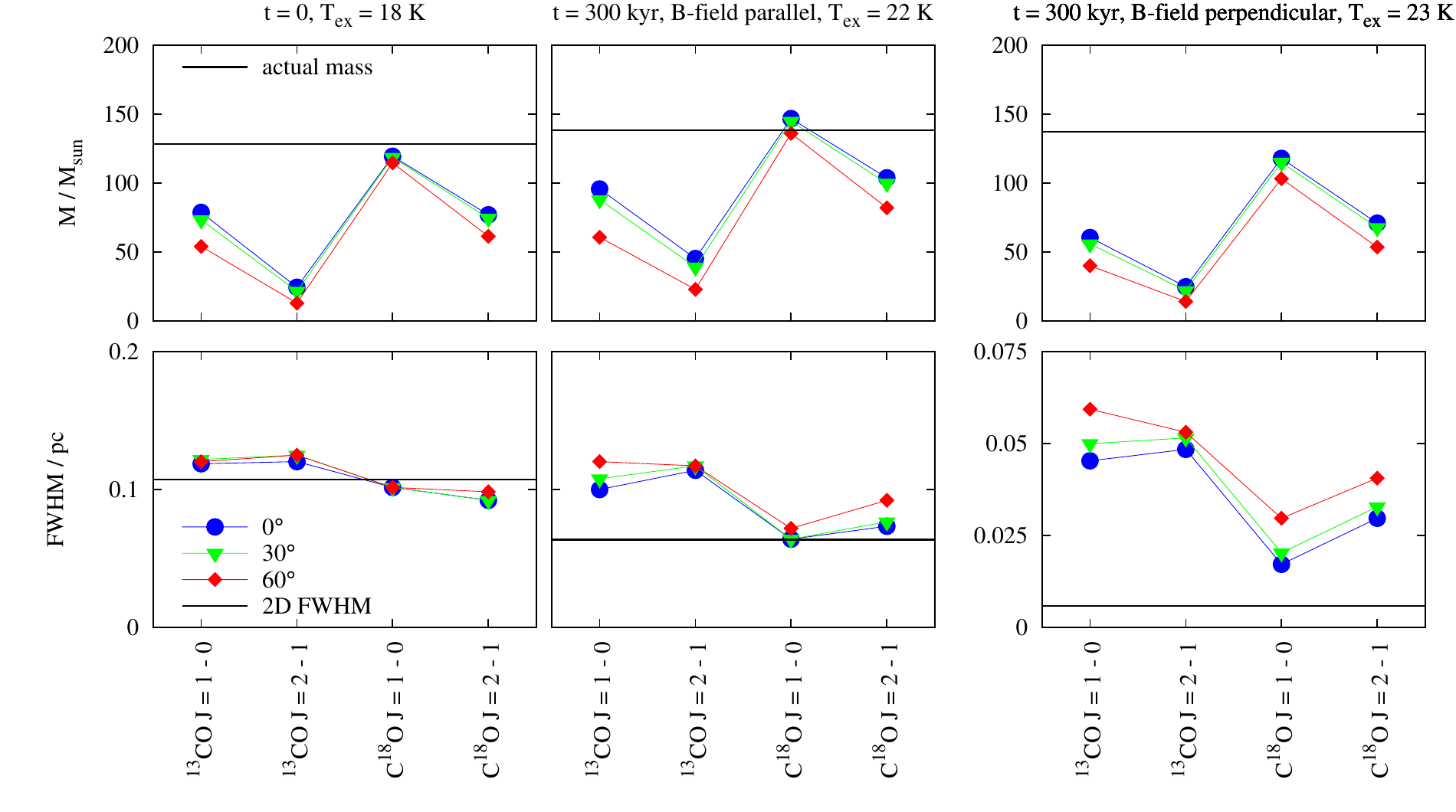}
 \caption{Impact of the inclination angle on the inferred masses (top row) and widths (bottom row) for the line emission of the runs with a CRIR of 1 $\times$ 10$^{-16}$ s$^{-1}$. Due to increasing optical thickness for increasing inclination, the obtained masses decrease and the FWHMs increase by up to a few 10\%. Note the different $y$-range in the bottom right panel.}
 \label{fig:incl} 
\end{figure*}

In order to test this hypothesis, we repeat the radiative transfer calculations for the simulations with a CRIR of 1 $\times$ 10$^{-16}$ s$^{-1}$, but now with an inclination of the main axis of the filaments of 30$^\circ$ and 60$^\circ$ with respect to the plane of the sky (0$^\circ$ means that the filament is in the plane of the sky, i.e. exactly perpendicular to the LOS, as assumed before). We plot the inferred masses and FWHMs in Fig.~\ref{fig:incl} for $t$ = 0 and for $t$ = 300 kyr for the parallel and perpendicular magnetic field case.

Overall, the inferred masses decrease only slightly with increasing inclination (top row of Fig.~\ref{fig:incl}), by a few percent for 30$^\circ$ and a few 10\% for 60$^\circ$. We attribute this to the fact that with increasing inclination the gas becomes increasingly optically thick, thus reducing the inferred mass.
The inferred FWHMs increase with increasing inclination (bottom row) although for $t$ = 0 the effect is only marginal. For the more condensed cases at $t$ = 300 kyr, the FWHMs for an inclination of 30$^\circ$ are larger by a few percent and for 60$^\circ$ larger by a few 10\%. Hence, just like the masses, the FWHMs are only marginally modified by inclination effects. In particular for 0.1 pc wide filaments (bottom left region of Fig.~\ref{fig:incl}), we find the inferred FWHMs to be accurate within a factor of $\sim$ 1.5 even for strongly inclined filaments.

For the values inferred from the dust emission (not shown), the effect of inclination is even less pronounced. Here we find that the inferred masses increase slightly with inclination by at most 10\% for the 60$^\circ$ case, but still fit the actual mass reasonably well. The FWHMs show (almost) no dependence on the inclination. For 30$^\circ$ no effect is measurable, and for 60$^\circ$ the FWHMs are only marginally smaller than for 0$^\circ$ by  1 -- 2\%.

Overall, it is thus possible that line emission maps produce slightly larger filament widths than the corresponding continuum emission maps when the filaments are inclined. However, given that the relative changes are rather small (a few 10\%) and that, for filaments in the plane of the sky, both line and dust emission give rather similar results (see Fig.~\ref{fig:CRIR_width}), we argue that in particular the large FWHMs found by various authors \citep{Juvela12a,Malinen12,Panopoulou14} cannot be attributed to inclination effects.

\section{Reliability of observable properties}
\label{sec:reliability}

\subsection{Filament widths}
\label{sec:widths_reliability}

As can be seen from Table~\ref{tab:models}, depending on the physical conditions like the CRIR or the magnetic field configuration, a wide range of actual filament widths can be obtained, even when starting from the same initial conditions \citep[see also Fig. 4 in][]{Seifried15}. In particular, more narrow filaments emerge in simulations with a low CRIR and/or a perpendicular magnetic field configuration, whereas high CRIRs and parallel field configurations induce an additional thermal/magnetic pressure which stabilizes the filaments against radial collapse \citep{Seifried15,Seifried16}. This wide range of FWHMs fits nicely with recent observational works, which report a rather larger range from a few times 0.01 pc, i.e. comparable to the FWHMs of some of our condensed filaments, up to about 1 pc \citep{Juvela12a,Malinen12,Panopoulou14,Sanchez14}.

\begin{figure*}
 \includegraphics[width=0.48\linewidth]{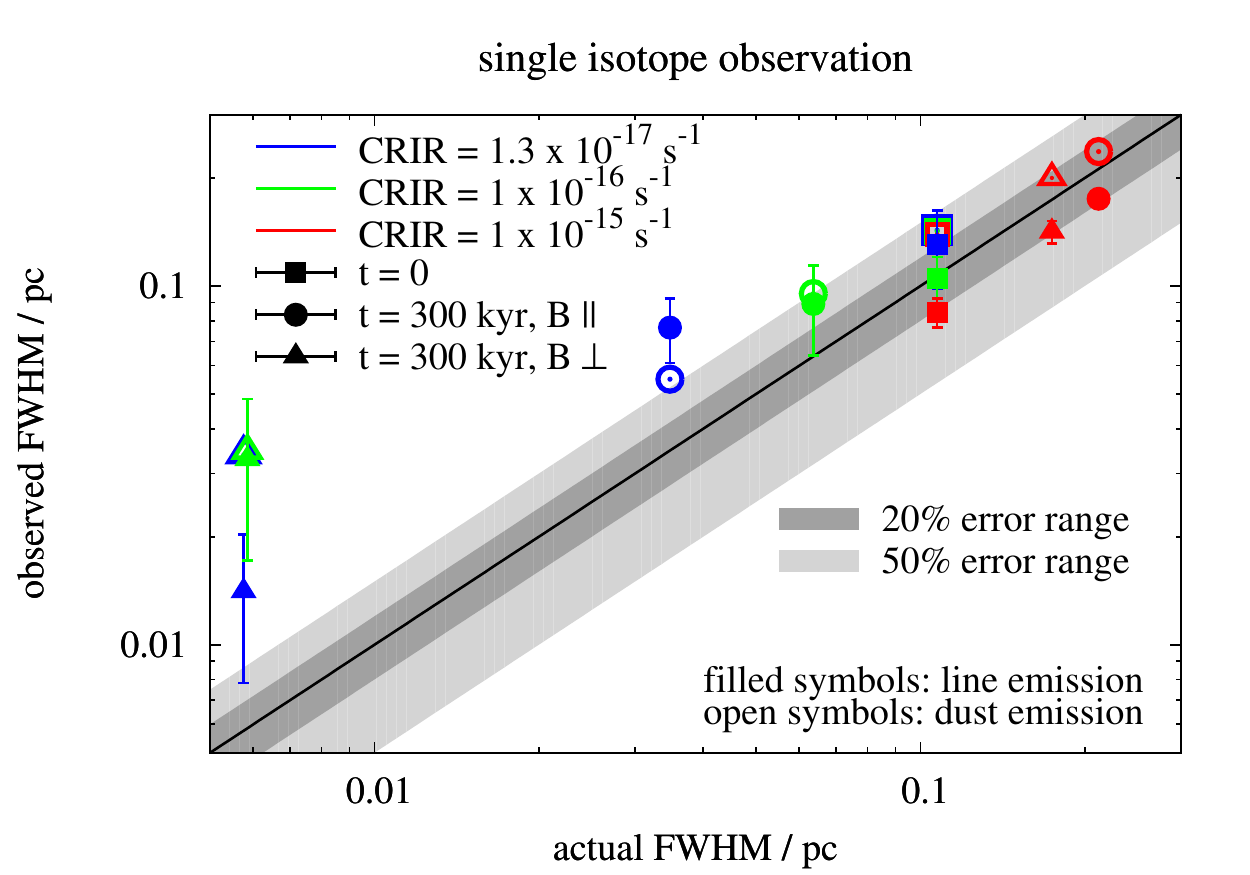}
  \includegraphics[width=0.48\linewidth]{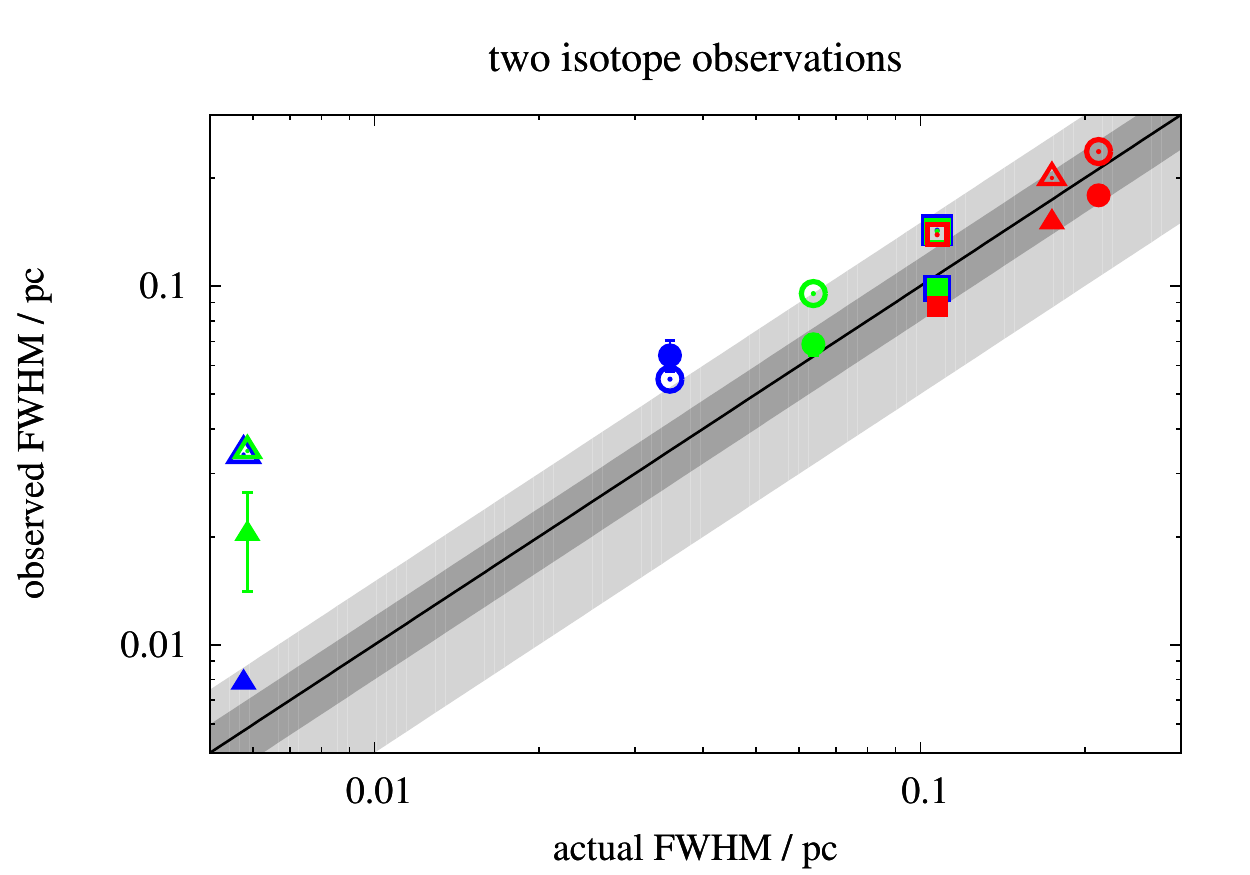}
 \caption{Reliability of observed filament widths in different situations like the time or magnetic field direction (different symbols) or different CRIRs (different colours). The reliability is assessed for the case of single isotope observations (left) and observations of both $^{13}$CO and C$^{18}$O (right). The filled symbols with errorbars combine the results of all four CO line transitions, open symbols display the results of the corresponding dust emission. For filaments with widths above 0.1 pc both line and dust emission observations can determine the FWHM relatively accurately indicated with the grey-shaded, relative error intervals. For very condensed filaments \mbox{(FWHMs $<$ 0.05 pc)}, the width is usually overestimated with relative errors above 50\%.}
 \label{fig:width_summary} 
\end{figure*}

In the left panel of Fig.~\ref{fig:width_summary} we plot the FWHMs extracted from the synthetic observations against the actual FWHM. The black line shows the one-to-one correspondence and the dark and light grey shaded areas indicate the 20\% and 50\% error, respectively. We summarise the results of the four CO line transition observations in one point (filled symbols), where the error bar indicates the range of obtained FWHMs. The corresponding dust emission result, obtained under the assumption of a beam size of 36'' and a source distance of 150 pc, is shown with an open symbol.

For filaments with an actual width above 0.1 pc, both line and dust emission observations can measure the actual FWHM with a high accuracy, i.e. with deviations $\leq$ 20\%. We note that this roughly corresponds to the uncertainty which is introduced by possible opacity and inclinations effects (Sections~\ref{sec:opacsingle} and~\ref{sec:incl}). Hence, we argue that observations finding large FWHMs can be considered as rather reliable.

For more narrow filaments, however, the obtained FWHMs become more inaccurate. For the very condensed filaments (FWHMs below a few times 0.01 pc; see also Table~\ref{tab:models}), the actual value is overestimated by more than 50\% and up to a factor of a few. As discussed in Sections~\ref{sec:temp} and~\ref{sec:subtherm}, we attribute this to the effect of the assumption of a fixed $T_\rmn{ex}$ and the optical depth combined with the subthermal excitation of CO, which reduce the derived column density in particular towards the centre of the filament. Our results thus indicate that very condensed filaments cannot be probed reliably with either line or continuum emission observations. The observations seem to rather result in a minimum threshold value for the FWHM of a few times 0.01 pc, which seems to fit nicely with aforementioned observational results \citep{Juvela12a,Malinen12,Sanchez14}.

\subsection{Filament masses and stability}
\label{sec:masses_reliability}

Throughout the paper we have used all wavelengths from 70 $\mu$m up to 2.6 mm to infer the column density and dust temperature from the continuum emission maps. We emphasise, however, that including e.g. only the \textit{Herschel} bands, i.e. 70, 160, 250, 350, and 500 $\mu$m, or excluding the 70 $\mu$m data, has only a marginal effect on the obtained mass, filament width, and central dust temperature. We attribute this to the fact that the intensities at 70 $\mu$m, 850 $\mu$m, 1.3 mm, and 2.6 mm are lower than at other wavelengths and thus have a negligible effect when fitting the modified blackbody spectrum to derive the column density and dust temperature of each pixel (Eq.~\ref{eq:BB}).

However, excluding all wavelengths $\geq$ 500 $\mu$m would in particular increase the inferred width by up to 30\%, the inferred mass is less affected. We attribute this to the fact that the densest parts are less reliably probed if the longer wavelengths are excluded \citep{Shetty09}. This results in lower central densities and thus -- since the outer, low density parts of the filament basically remain unchanged -- larger \textit{apparent} filament widths. Hence, overall our results suggest that the masses obtained via continuum emission show a high accuracy (within a few percent, see Fig.~\ref{fig:CRIR_mass}) for the various wavelength regimes considered. We therefore suggest to use continuum observations to infer the mass of star-forming filaments. We note, however, that a network of overlapping filaments as seen in observations might complicate the determination of the filament mass and stability \citep[e.g.][]{Andre10,Busquet13}.

In contrast to that, the masses obtained via line emission observations show partly significant deviations of up to a factor of 10 from the actual mass. Furthermore, even when observing lines of two isotopes at the same time, a reliable calculation of $T_\rmn{ex}$ (and thus $T_\rmn{gas}$) is often hardly possible due to different spectral shapes (see Section~\ref{sec:opacmulti}). For this reason, in observations \citep[e.g.][]{Andre10,Arzoumanian11,Palmeirim13,Kainulainen13,Kainulainen16b} often $T_\rmn{dust}$ or a freely assumed temperature (motivated by a typical value for $T_\rmn{dust}$) is used  in order to assess the stability of the filaments via comparing their mass-per-unit-length to the critical value \citep[Eq.~\ref{eq:crit}, but see also][]{Ostriker64}. However, as shown in Fig.~\ref{fig:temp}, $T_\rmn{gas}$ does not necessarily resemble $T_\rmn{dust}$, which can be lower by up to a factor of a few. Hence, when using the inferred \textit{dust} instead of the \textit{gas} temperature in Eq.~\ref{eq:crit}, this might lead to wrong conclusions about the stability of star-forming filaments. In particular, filaments might appear significantly more gravitationally unstable than they actually are.

\subsection{Observations of two isotopes}
\label{sec:opacmulti}

So far, we have assumed that we have observed line emission of a \textit{single} CO isotope only. In case that observations of \textit{two} isotopes are available, the combination of both can be used to correct for the optical depth \citep[e.g.][]{Myers83}. This is usually done on a channel-by-channel basis. However, comparing the corresponding spectra of $^{13}$CO and C$^{18}$O for one of our simulated filaments (Fig.~\ref{fig:spectra}), we find that they do not have the same functional (Gaussian) shape. This happens when the variation of $\tau$ with the channel velocity is different for different isotopes. Hence, in this case we do not recommend to correct for the optical depth on a channel-by-channel basis but rather by using the integrated intensity \citep[see also][]{Arzoumanian13}. We emphasise, however, that this prevents us from determining $T_\rmn{ex}$, which thus remains a free parameter.\footnote{This is due to the fact that from Eq.~\ref{eq:tau_iso} we can only determine a ``channel-averaged'' $\tau$, whereas Eq.~\ref{eq:tau} -- from which $T_\rmn{ex}$ could be inferred -- is on an individual channel basis.}

\begin{figure*}
 \includegraphics[width=0.48\linewidth]{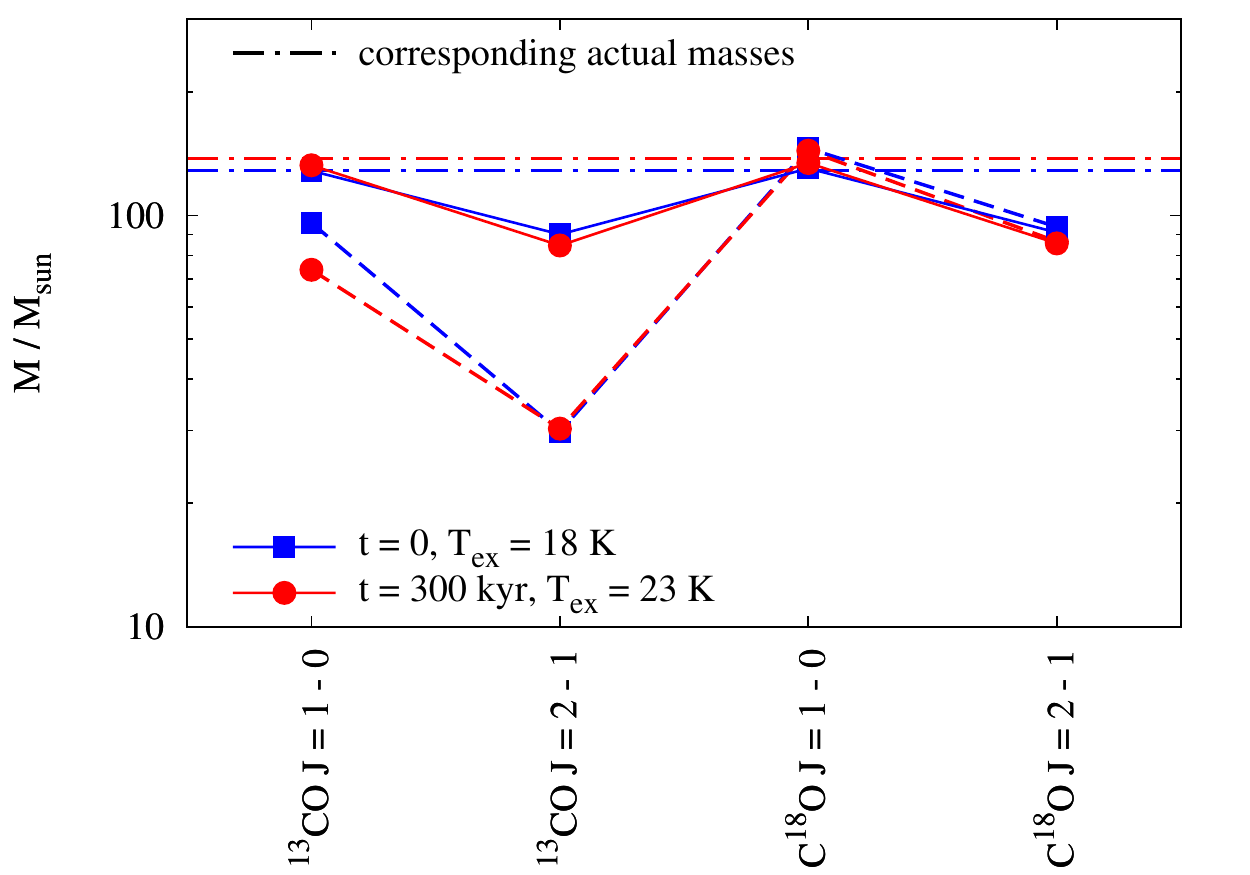}
 \includegraphics[width=0.48\linewidth]{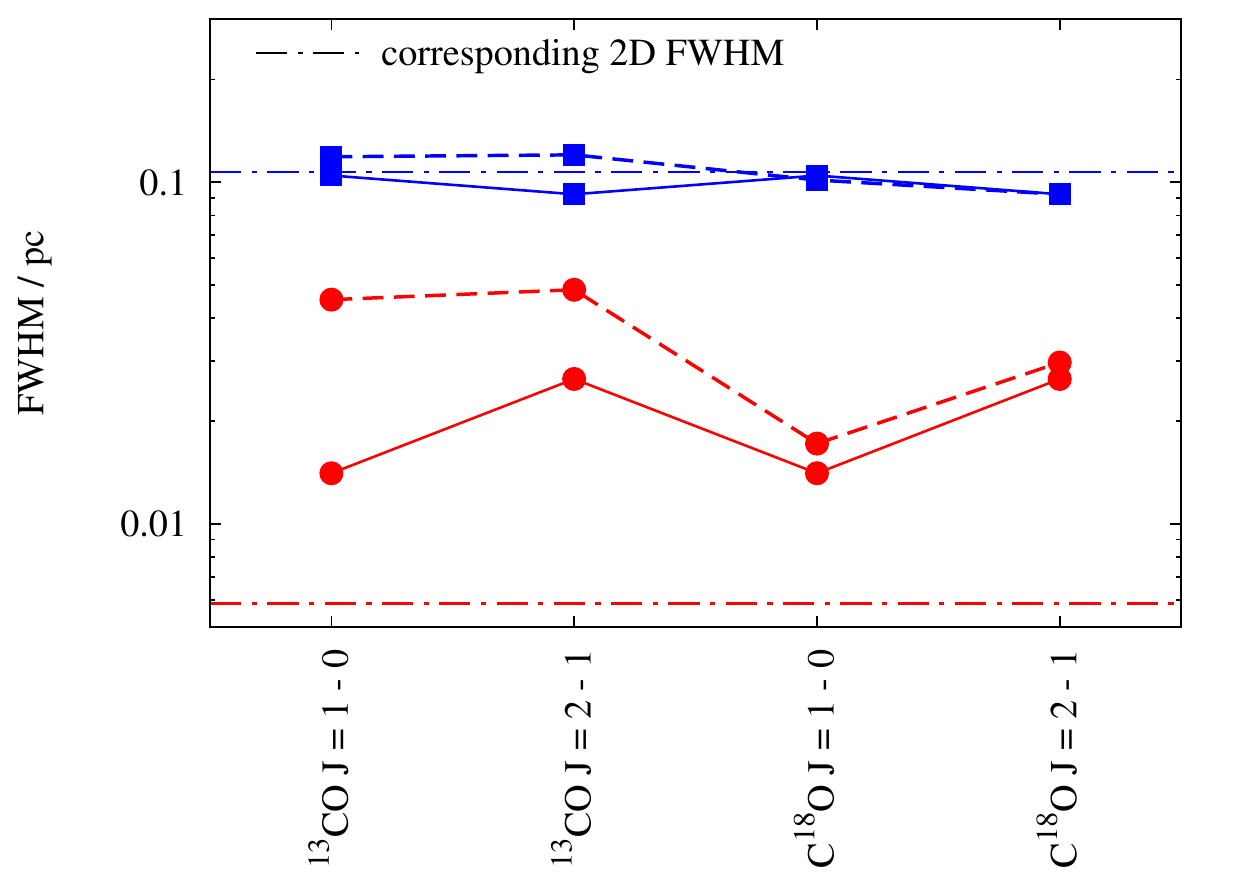}
 \caption{Comparison of masses (left) and FWHMs (right) obtained from single transition observations with opacity correction (dashed lines, Eq.~\ref{eq:NCO}) and that combining the observations of the isotopes $^{13}$CO and C$^{18}$O to correct for the optical depth (solid lines, Eq.~\ref{eq:NCO_opac}). The considered run has a CRIR of 1 $\times$ 10$^{-16}$ s$^{-1}$ and a perpendicular magnetic field. For $t$ = 0 and 300 kyr we assumed $T_\rmn{ex}$ = 18 and \mbox{23 K}, respectively. The straight lines give the actual mass and FWHM of the filament.}
 \label{fig:mass_opac}
\end{figure*}

Using the integrated intensity, the optical depth $\tau_1$ of isotope 1 is then obtained by solving
\begin{equation}
 \frac{\int T_\rmn{B,1} \rmn{d}v}{\int T_\rmn{B,2} \rmn{d}v} = \frac{1 - \rmn{e}^{-\tau_\rmn{1}}}{1 - \rmn{e}^{-R \tau_\rmn{1}}}
 \label{eq:tau_iso}
\end{equation}
for $\tau_\rmn{1}$, where $R$ is the isotope abundance ratio. The optical depth $\tau_2$ of the isotope 2 is then given by $R \times \tau_\rmn{1}$. Next, Eq.~\ref{eq:NCO} can be rearranged to yield the total CO column density using the afore determined $\tau$ as well as a given excitation temperature:
\begin{flalign}
  & N_\rmn{CO} =  \label{eq:NCO_opac} \\
   & \frac{8 \pi \nu^3}{c^3} \frac{1}{A} \, f(T_\rmn{ex}) \, \frac{Q}{g_\rmn{u}} \, \frac{k_\rmn{B}/(h \nu)}{ f(T_\rmn{ex}) - f(T_\rmn{bg}) } \, \frac{\tau}{1-\rmn{e}^{-\tau}} \, \rmn{e}^{\frac{E_\rmn{u}}{T_\rmn{ex}}} \int T_\rmn{B} \rmn{d}v \, . \nonumber
\end{flalign}

In the following we assume that we have $^{13}$CO and C$^{18}$O observations of the same $J$ transition available. We consider the optical depth effect for the run with a perpendicular magnetic field and a CRIR of 1 $\times$ 10$^{-16}$ s$^{-1}$ for $t$ = 0 and 300 kyr in Fig.~\ref{fig:mass_opac}. In this run the FWHM decreases from about 0.1 pc to 0.0059 pc over time, which makes it an interesting target. We compare the masses and FWHMs obtained via single isotope observations (dashed lines) and those obtained using both isotopes (solid lines). We constrain ourselves to the excitation temperatures of $T_\rmn{ex}$ = 18 K for $t$ = 0 and $T_\rmn{ex}$ = 23 K for $t$ = 300 kyr used already for the single isotope case\footnote{For other temperatures no significant differences occur.}.

The masses are now matched significantly better than for the single transition observations, which previously showed differences up to a factor of $\sim$ 4, although still deviations of up to $\sim$ 40\% occur. However, both $J$ = 2 -- 1 transitions still systematically underestimate the masses. As discussed in Sections~\ref{sec:temp} and~\ref{sec:subtherm}, we attribute this to the variation of the temperature along the LOS as well as the sub-thermal excitation of CO in the outer parts.

Furthermore, using two isotopes to correct for $\tau$ seems also to slightly improve the ability to probe rather condensed filaments. For $t$ = 300 kyr the match is now somewhat better, although the actual FWHM is still overestimated by up to a factor of $\sim$ 5. For $t$ = 0 the FWHM is very well matched with deviations of about 10\% --  as for the single transition observations. We summarise the reliability of the obtained FWHMs in the right panel of Fig.~\ref{fig:width_summary} for all snapshots with $G_0 = 1.7$. Overall, we find that the variations between different lines for a single snapshot (actual FWHM) are now significantly reduced as indicated by the smaller errorbars compared to the single line observations. Furthermore, for the condensed filaments the ability to accurately measure the FWHM is somewhat improved.

Inspecting the values of $\tau$ (not shown), for $J$ = 1 -- 0 we obtain typical values of 1 -- 5 for $^{13}$CO and 0.1 -- 0.5 for C$^{18}$O. For $J$ = 2 -- 1, the values of $\tau$ are on average only slightly higher reaching up to at most 10 and 1 for $^{13}$CO and C$^{18}$O, respectively. Hence, for both transitions the values are in very good agreement with those values obtained from our single transition observations as well as actual observations by \citet{Arzoumanian13}. 

To summarise, for observations it seems desirable to obtain -- if possible -- observations of two isotopes in order to improve the accuracy of observational parameters like filament mass and width. If the spectra have different shapes, we do not recommend to make the opacity correction on a channel-by-channel basis.

\subsection{Caveats}
\label{sec:caveats}

In the simulations the effect of freeze-out is taken into account in a simplified manner by reducing the overall abundance of C and O. This approach, however, does not take into account any density or CR dependence. In order to test the effect of CO freeze-out in more detail, we incorporated a post-processing step following the approach described in detail in \citet{Glover16}. Before running the radiative transfer calculations, we update the CO number density by multiplying it with a factor of
\begin{equation}
 f_\rmn{CO} = \frac{k_{CR}}{k_{CR} + k_\rmn{ads}} \, ,
 \label{eq:freezeout}
\end{equation}
which gives the fraction of CO left over in the gas phase. Here,
\begin{equation}
 k_\rmn{CR} = 5.7 \times 10^{-13} \times \frac{\text{CRIR}}{10^{-17} \text{s$^{-1}$}} \text{\, s$^{-1}$ molecule$^{-1}$}
 \label{eq:CRdesorption}
\end{equation}
is the CR induced desorption rate of CO \citep{Herbst06}, and 
\begin{equation}
 k_\rmn{ads} = 3.44 \times 10^{-18} \sqrt{T_\rmn{gas}} \, n_\rmn{H,tot} \text{\, s$^{-1}$ molecule$^{-1}$}
\end{equation}
the adsorption rate due to collisions of CO with the dust grains \citep{Hollenbach09}, where $n_\rmn{H,tot}$ is the hydrogen nuclei number density. We note that since for $T_\rmn{dust} <$ 20 K as present in the simulations (see Fig.~\ref{fig:dusttemp}), the desorption process is dominated by CRs, we here omitted the process of thermal desorption of CO.

\begin{figure*}
 \includegraphics[width=0.48\textwidth]{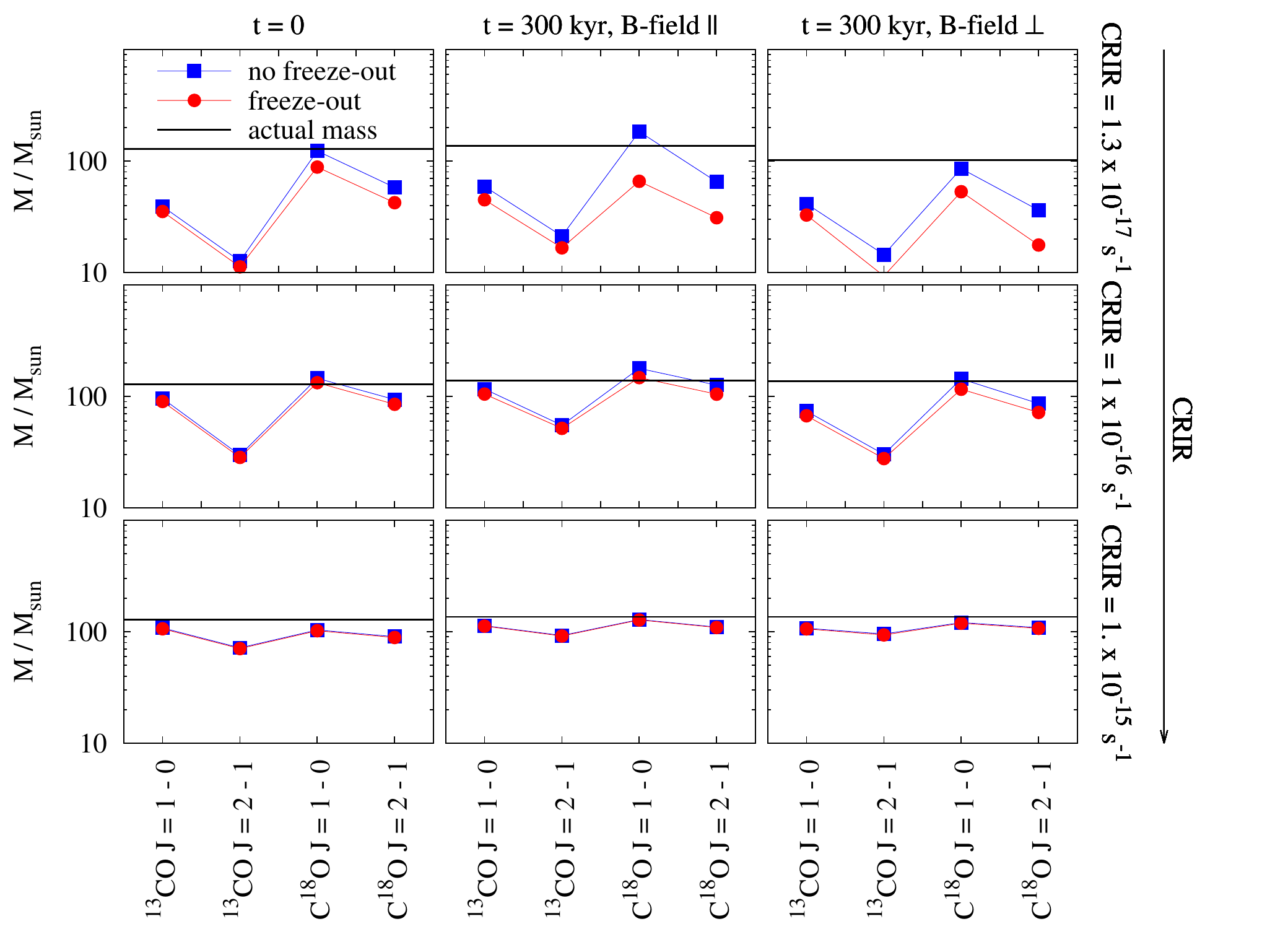}
 \includegraphics[width=0.48\textwidth]{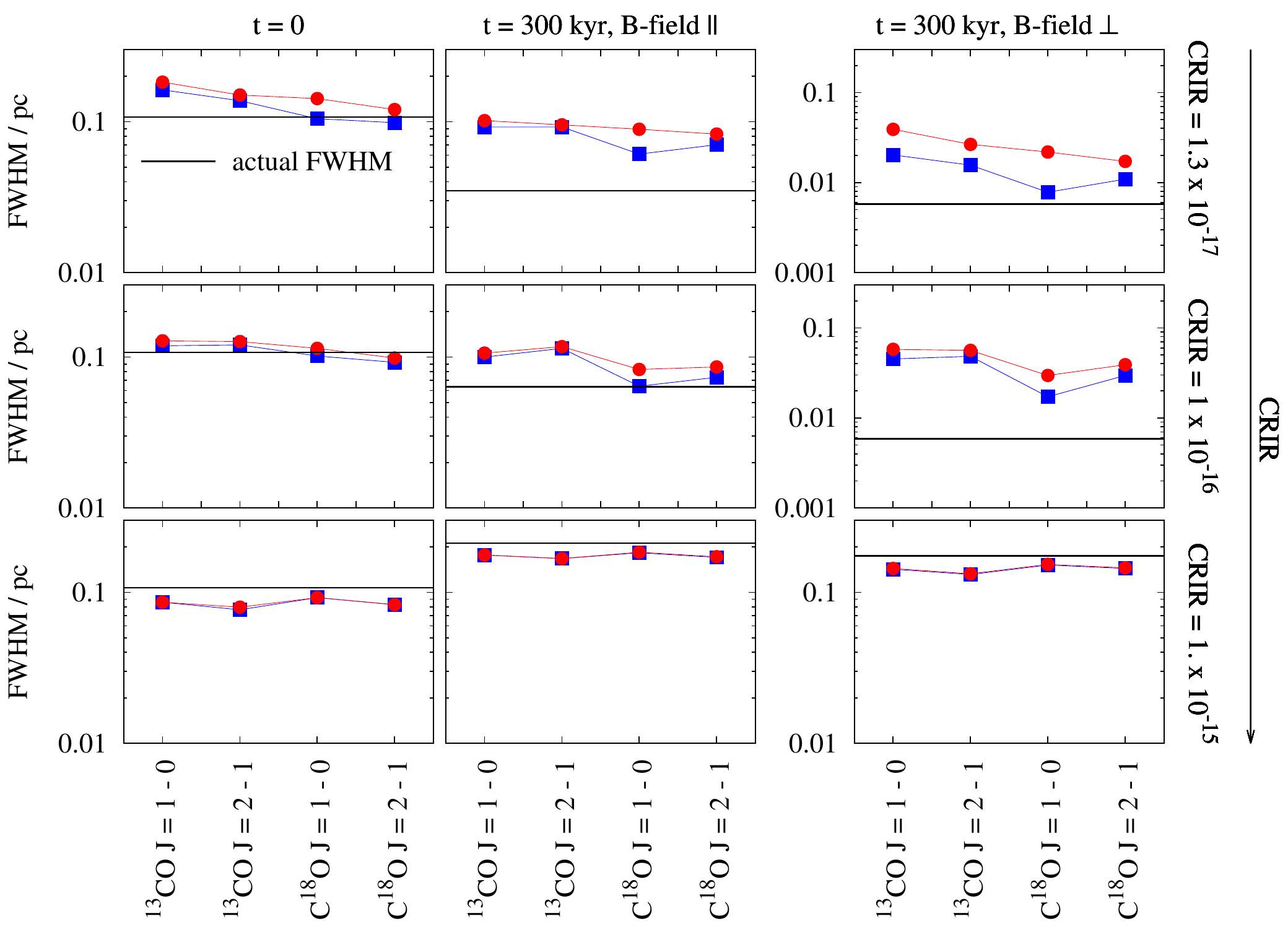}
 \caption{Effect of incorporating a simple CO freeze-out approximation (red lines) before the radiative transfer calculations on the inferred masses (left) and FWHMs (right). Overall, the masses decrease and the FWHMs increase compared to those without the approximation (blue lines). CO freeze-out thus intensifies the problem of overestimating the width of condensed filaments.}
 \label{fig:freeze}
\end{figure*}

The effect of CO freeze-out concerning the inferred mass and FWHM is shown in Fig.~\ref{fig:freeze} using the default (lower) excitation temperatures given in Fig.~\ref{fig:CRIR_mass}. In general, freeze-out results in somewhat lower masses and larger FWHMs compared to the results without the aforementioned post-processing step. This is due to the reduction of CO in particular in the center of the filaments, where the density is high. As expected from Eqs.~\ref{eq:freezeout} and~\ref{eq:CRdesorption}, the effect decreases with increasing CRIR, and is almost negligible for a CRIR of \mbox{$1 \times 10^{-15}$ s$^{-1}$}. However, even for lower CRIRs the effect of CO freeze-out is moderate for both mass and FWHM (a few 10\% at most). The only exception occurs at $t$ = 300 kyr for the run with a perpendicular magnetic field and a CRIR of \mbox{$1.3 \times 10^{-17}$ s$^{-1}$} (top right panel). Here, freeze-out leads to an increase of the inferred FWHM by up to a factor of 3. Hence, overall CO freeze-out intensifies the afore described problem of overestimating the width of condensed filaments, thus supporting the findings of this paper.

Furthermore, we note that the imperfect thermal coupling of gas and dust even in the center (Fig.~\ref{fig:temp}) could be in parts a consequence of the choice of our gas-dust energy transfer rate due to collision between gas and dust, which is based on \citet{Goldsmith01}. In order to test this, we repeated the run with a parallel magnetic field and a CRIR of $1 \times 10^{-16}$ s$^{-1}$ with a 6 times higher energy transfer rate as used by \citet{Krumholz11} \citep[see also][]{Hollenbach89}. We find that the thermal coupling is improved but still not perfect. The gas temperature in the inner 0.1 pc of the filament drops by up to 8 K compared to the run with a lower energy transfer rate, whereas the dust temperature remains almost unchanged. Hence, in the center $T_\rmn{gas}$ is still about 3 -- 10 K above $T_\rmn{dust}$. In the outer parts no measurable changes for both $T_\rmn{gas}$ and $T_\rmn{dust}$ occur. Overall, we thus find that the qualitative behaviour of the temperature evolution does not change with a higher gas-dust energy transfer rate, and that gas and dust are still not (yet) tightly thermally coupled.

Finally, we note that although the dust remains optically thin, its own thermal radiation can contribute to its heating. Since we do not account for this effect, in the center of the filament, where the ISRF is strongly attenuated, we thus might slightly underestimate $T_\rmn{dust}$.

\section{Conclusions}
\label{sec:conclusion}

We present synthetic emission maps of $^{13}$CO, C$^{18}$O, and dust of simulations of star-forming filaments including one of the largest chemical networks (37 species and 287 reactions) used to date in 3D-MHD simulations \citep{Seifried16}. From the synthetic observations, we calculate filament masses and widths, which we compare to the actual values as extracted from the simulation data. This allows us to test their reliability and to investigate the impact of the cosmic ray ionisation rate (CRIR) and the magnetic field structure. Our main conclusions are:
\begin{itemize}
 \item For filaments with widths around 0.1 pc or above, both line and dust emission observations can be used to determine the full-width-half-maximum (FWHM) with a high accuracy (deviations $\leq$ 20\%). For filaments with an actual FWHM $<$ 0.1 pc, the FWHMs derived from synthetic line and dust emission maps and the actual ones diverge as the synthetic observations consistently overestimate the width by up to a factor of a few.
 \item Masses obtained from the dust emission maps reproduce the actual masses very accurately to within a few percent. For this reason, we suggest to use these to analyse the stability of filaments.
 \item In contrast, masses inferred from line emission observations show strong variations by up to a factor of 10 between the individual lines, and partly clear deviations from the actual mass. For both $^{13}$CO and C$^{18}$O, the $J$  = 2 -- 1 lines significantly underestimate the actual masses.
\end{itemize}

The results reported here suggest a \textit{minimum detectable FWHM} of a few times 0.01 pc for line emission maps and continuum observations with resolutions comparable to \textit{Herschel}, in good agreement with minimal values obtained by recent observations. The detection of such condensed filaments possibly indicates a low CRIR and/or a perpendicular magnetic field in the filament. On the other hand, a high CRIR or a parallel field induces an additional thermal/magnetic pressure, which stabilizes the filaments against radial collapse resulting in widths $\geq$ 0.1 pc.

We attribute the inaccurate estimate of filament masses and widths of narrow filaments using molecular line observations to
\begin{itemize}
 \item strong variations of the gas temperature $T_\rmn{gas}$ along the LOS due to accretion shocks in the filaments, which are not accounted for by using a constant excitation temperature, and
 \item self-absorption features in some of the spectra due to the high optical depth $\tau$ and the fact that CO is only subthermally excited.
\end{itemize}

Furthermore, we emphasise that $T_\rmn{gas}$ and $T_\rmn{dust}$ do in general not match each other but can differ by a factor of a few. This would significantly affect the stability analysis of filaments when using $T_\rmn{dust}$. We find that the observed dust temperatures $T_\rmn{dust}$ slightly overestimate the actual ones by a few K \citep{Shetty09}, whereas the polytropic relation between $T_\rmn{dust}$ and the gas density can be probed reliably \citep[compare e.g.][]{Arzoumanian11}.

We also show that inclination effects increase the mass and width obtained from line emission maps only slightly by a few 10\%. Values obtained from dust emission maps are almost not affected. A similar uncertainty is induced by the optical depth correction in the line emission observations: the observed lines have optical depths of $\tau$ = 1 -- 5 for $^{13}$CO and $\tau$ = 0.1 -- 1 for C$^{18}$O in good agreement with observations. Hence, when not correcting for $\tau$, the observed masses would decrease and the FWHMs increase by a few 10\%.

Finally, we show that the reliability of the obtained mass and width is improved when using two isotopes to correct for the optical depth, although the width of narrow filaments is still overestimated. However, in case the shapes of the two spectra are different, we do not recommend to do this on a channel-by-channel basis but rather to use the 0th moment maps.

\section*{Acknowledgements}

The authors thank the referee for his very helpful report, which contributed to improve the quality of the paper. The authors acknowledge funding by the Deutsche Forschungsgemeinschaft (DFG) via the Sonderforschungsbereich SFB 956 \textit{Conditions and Impact of Star Formation} (subprojects  A4, A6, C5) and the Schwerpunktprogramm SPP 1573 \textit{Physics of the ISM}. Furthermore, DS, S. Suri, and SW acknowledge funding by the Bonn-Cologne Graduate School. SW also acknowledges funding by the ERC Starting Grant No. 679852 ``RADFEEDBACK''. The simulations were performed on JURECA at the Computing Center J\"ulich. The FLASH code was developed partly by the DOE-supported Alliances Center for Astrophysical Thermonuclear Flashes (ASC) at the University of Chicago.

\label{lastpage}


\begin{thebibliography}{73}
\expandafter\ifx\csname natexlab\endcsname\relax\def\natexlab#1{#1}\fi

\bibitem[{{Andr{\'e}} {et~al.}(2010){Andr{\'e}}, {Men'shchikov}, {Bontemps},
  {K{\"o}nyves}, {Motte}, {Schneider}, {Didelon}, {Minier}, {Saraceno},
  {Ward-Thompson}, {di Francesco}, {White}, {Molinari}, {Testi}, {Abergel},
  {Griffin}, {Henning}, {Royer}, {Mer{\'{\i}}n}, {Vavrek}, {Attard},
  {Arzoumanian}, {Wilson}, {Ade}, {Aussel}, {Baluteau}, {Benedettini},
  {Bernard}, {Blommaert}, {Cambr{\'e}sy}, {Cox}, {di Giorgio}, {Hargrave},
  {Hennemann}, {Huang}, {Kirk}, {Krause}, {Launhardt}, {Leeks}, {Le Pennec},
  {Li}, {Martin}, {Maury}, {Olofsson}, {Omont}, {Peretto}, {Pezzuto}, {Prusti},
  {Roussel}, {Russeil}, {Sauvage}, {Sibthorpe}, {Sicilia-Aguilar}, {Spinoglio},
  {Waelkens}, {Woodcraft}, \& {Zavagno}}]{Andre10}
{Andr{\'e}}, P., {Men'shchikov}, A., {Bontemps}, S., {et~al.} 2010, \aap, 518,
  L102

\bibitem[{{Arzoumanian} {et~al.}(2011){Arzoumanian}, {Andr{\'e}}, {Didelon},
  {K{\"o}nyves}, {Schneider}, {Men'shchikov}, {Sousbie}, {Zavagno}, {Bontemps},
  {di Francesco}, {Griffin}, {Hennemann}, {Hill}, {Kirk}, {Martin}, {Minier},
  {Molinari}, {Motte}, {Peretto}, {Pezzuto}, {Spinoglio}, {Ward-Thompson},
  {White}, \& {Wilson}}]{Arzoumanian11}
{Arzoumanian}, D., {Andr{\'e}}, P., {Didelon}, P., {et~al.} 2011, \aap, 529, L6

\bibitem[{{Arzoumanian} {et~al.}(2013){Arzoumanian}, {Andr{\'e}}, {Peretto}, \&
  {K{\"o}nyves}}]{Arzoumanian13}
{Arzoumanian}, D., {Andr{\'e}}, P., {Peretto}, N., \& {K{\"o}nyves}, V. 2013,
  \aap, 553, A119

\bibitem[{{Bally}(1986)}]{Bally86}
{Bally}, J. 1986, Science, 232, 185

\bibitem[{{Bally} {et~al.}(1987){Bally}, {Langer}, {Stark}, \&
  {Wilson}}]{Bally87}
{Bally}, J., {Langer}, W.~D., {Stark}, A.~A., \& {Wilson}, R.~W. 1987, \apjl,
  312, L45

\bibitem[{{Busquet} {et~al.}(2013){Busquet}, {Zhang}, {Palau}, {Liu},
  {S{\'a}nchez-Monge}, {Estalella}, {Ho}, {de Gregorio-Monsalvo}, {Pillai},
  {Wyrowski}, {Girart}, {Santos}, \& {Franco}}]{Busquet13}
{Busquet}, G., {Zhang}, Q., {Palau}, A., {et~al.} 2013, \apjl, 764, L26

\bibitem[{{Caselli} {et~al.}(1998){Caselli}, {Walmsley}, {Terzieva}, \&
  {Herbst}}]{Caselli98}
{Caselli}, P., {Walmsley}, C.~M., {Terzieva}, R., \& {Herbst}, E. 1998, \apj,
  499, 234

\bibitem[{{Ceccarelli} {et~al.}(2011){Ceccarelli}, {Hily-Blant}, {Montmerle},
  {Dubus}, {Gallant}, \& {Fiasson}}]{Ceccarelli11}
{Ceccarelli}, C., {Hily-Blant}, P., {Montmerle}, T., {et~al.} 2011, \apjl, 740,
  L4

\bibitem[{{Chapman} {et~al.}(2011){Chapman}, {Goldsmith}, {Pineda}, {Clemens},
  {Li}, \& {Kr{\v c}o}}]{Chapman11}
{Chapman}, N.~L., {Goldsmith}, P.~F., {Pineda}, J.~L., {et~al.} 2011, \apj,
  741, 21

\bibitem[{{Clark} {et~al.}(2012){Clark}, {Glover}, \& {Klessen}}]{Clark12}
{Clark}, P.~C., {Glover}, S.~C.~O., \& {Klessen}, R.~S. 2012, \mnras, 420, 745

\bibitem[{{Clark} {et~al.}(2013){Clark}, {Glover}, {Ragan}, {Shetty}, \&
  {Klessen}}]{Clark13}
{Clark}, P.~C., {Glover}, S.~C.~O., {Ragan}, S.~E., {Shetty}, R., \& {Klessen},
  R.~S. 2013, \apjl, 768, L34

\bibitem[{{Dobbs} {et~al.}(2008){Dobbs}, {Glover}, {Clark}, \&
  {Klessen}}]{Dobbs08b}
{Dobbs}, C.~L., {Glover}, S.~C.~O., {Clark}, P.~C., \& {Klessen}, R.~S. 2008,
  \mnras, 389, 1097

\bibitem[{{Draine}(1978)}]{Draine78}
{Draine}, B.~T. 1978, \apjs, 36, 595

\bibitem[{{Dubey} {et~al.}(2008){Dubey}, {Fisher}, {Graziani}, {Jordan},
  {Lamb}, {Reid}, {Rich}, {Sheeler}, {Townsley}, \& {Weide}}]{Dubey08}
{Dubey}, A., {Fisher}, R., {Graziani}, C., {et~al.} 2008, in Astronomical
  Society of the Pacific Conference Series, Vol. 385, Numerical Modeling of
  Space Plasma Flows, ed. N.~V. {Pogorelov}, E.~{Audit}, \& G.~P. {Zank}, 145

\bibitem[{{Dullemond}(2012)}]{Dullemond12}
{Dullemond}, C.~P. 2012, {RADMC-3D: A multi-purpose radiative transfer tool},
  astrophysics Source Code Library

\bibitem[{{Federrath} {et~al.}(2010){Federrath}, {Banerjee}, {Clark}, \&
  {Klessen}}]{Federrath10}
{Federrath}, C., {Banerjee}, R., {Clark}, P.~C., \& {Klessen}, R.~S. 2010,
  \apj, 713, 269

\bibitem[{{Flower} {et~al.}(2005){Flower}, {Pineau Des For{\^e}ts}, \&
  {Walmsley}}]{Flower05}
{Flower}, D.~R., {Pineau Des For{\^e}ts}, G., \& {Walmsley}, C.~M. 2005, \aap,
  436, 933

\bibitem[{{Friesen} {et~al.}(2016){Friesen}, {Bourke}, {Di Francesco},
  {Gutermuth}, \& {Myers}}]{Friesen16}
{Friesen}, R.~K., {Bourke}, T.~L., {Di Francesco}, J., {Gutermuth}, R., \&
  {Myers}, P.~C. 2016, \apj, 833, 204

\bibitem[{{Fryxell} {et~al.}(2000){Fryxell}, {Olson}, {Ricker}, {Timmes},
  {Zingale}, {Lamb}, {MacNeice}, {Rosner}, {Truran}, \& {Tufo}}]{Fryxell00}
{Fryxell}, B., {Olson}, K., {Ricker}, P., {et~al.} 2000, \apjs, 131, 273

\bibitem[{{Glover} \& {Clark}(2012)}]{Glover12}
{Glover}, S.~C.~O. \& {Clark}, P.~C. 2012, \mnras, 421, 116

\bibitem[{{Glover} \& {Clark}(2016)}]{Glover16}
{Glover}, S.~C.~O. \& {Clark}, P.~C. 2016, \mnras, 456, 3596

\bibitem[{{Glover} {et~al.}(2010){Glover}, {Federrath}, {Mac Low}, \&
  {Klessen}}]{Glover10}
{Glover}, S.~C.~O., {Federrath}, C., {Mac Low}, M.-M., \& {Klessen}, R.~S.
  2010, \mnras, 404, 2

\bibitem[{{Glover} \& {Mac Low}(2007)}]{Glover07}
{Glover}, S.~C.~O. \& {Mac Low}, M.-M. 2007, \apj, 659, 1317

\bibitem[{{Goldsmith}(2001)}]{Goldsmith01}
{Goldsmith}, P.~F. 2001, \apj, 557, 736

\bibitem[{{Grassi} {et~al.}(2014){Grassi}, {Bovino}, {Schleicher}, {Prieto},
  {Seifried}, {Simoncini}, \& {Gianturco}}]{Grassi14}
{Grassi}, T., {Bovino}, S., {Schleicher}, D.~R.~G., {et~al.} 2014, \mnras, 439,
  2386

\bibitem[{{Hacar} {et~al.}(2013){Hacar}, {Tafalla}, {Kauffmann}, \&
  {Kov{\'a}cs}}]{Hacar13}
{Hacar}, A., {Tafalla}, M., {Kauffmann}, J., \& {Kov{\'a}cs}, A. 2013, \aap,
  554, A55

\bibitem[{{Herbst} \& {Cuppen}(2006)}]{Herbst06}
{Herbst}, E. \& {Cuppen}, H.~M. 2006, Proceedings of the National Academy of
  Science, 103, 12257

\bibitem[{{Hernandez} \& {Tan}(2011)}]{Hernandez11}
{Hernandez}, A.~K. \& {Tan}, J.~C. 2011, \apj, 730, 44

\bibitem[{{Hildebrand}(1983)}]{Hildebrand83}
{Hildebrand}, R.~H. 1983, \qjras, 24, 267

\bibitem[{{Hocuk} {et~al.}(2016){Hocuk}, {Cazaux}, {Spaans}, \&
  {Caselli}}]{Hocuk16}
{Hocuk}, S., {Cazaux}, S., {Spaans}, M., \& {Caselli}, P. 2016, \mnras, 456,
  2586

\bibitem[{{Hollenbach} {et~al.}(2009){Hollenbach}, {Kaufman}, {Bergin}, \&
  {Melnick}}]{Hollenbach09}
{Hollenbach}, D., {Kaufman}, M.~J., {Bergin}, E.~A., \& {Melnick}, G.~J. 2009,
  \apj, 690, 1497

\bibitem[{{Hollenbach} \& {McKee}(1989)}]{Hollenbach89}
{Hollenbach}, D. \& {McKee}, C.~F. 1989, \apj, 342, 306

\bibitem[{{Inoue} \& {Inutsuka}(2012)}]{Inoue12}
{Inoue}, T. \& {Inutsuka}, S.-i. 2012, \apj, 759, 35

\bibitem[{{Juvela} {et~al.}(2012{\natexlab{a}}){Juvela}, {Malinen}, \&
  {Lunttila}}]{Juvela12b}
{Juvela}, M., {Malinen}, J., \& {Lunttila}, T. 2012{\natexlab{a}}, \aap, 544,
  A141

\bibitem[{{Juvela} {et~al.}(2012{\natexlab{b}}){Juvela}, {Ristorcelli},
  {Pagani}, {Doi}, {Pelkonen}, {Marshall}, {Bernard}, {Falgarone}, {Malinen},
  {Marton}, {McGehee}, {Montier}, {Motte}, {Paladini}, {T{\'o}th}, {Ysard},
  {Zahorecz}, \& {Zavagno}}]{Juvela12a}
{Juvela}, M., {Ristorcelli}, I., {Pagani}, L., {et~al.} 2012{\natexlab{b}},
  \aap, 541, A12

\bibitem[{{Kainulainen} {et~al.}(2016{\natexlab{a}}){Kainulainen}, {Hacar},
  {Alves}, {Beuther}, {Bouy}, \& {Tafalla}}]{Kainulainen16a}
{Kainulainen}, J., {Hacar}, A., {Alves}, J., {et~al.} 2016{\natexlab{a}}, \aap,
  586, A27

\bibitem[{{Kainulainen} {et~al.}(2013){Kainulainen}, {Ragan}, {Henning}, \&
  {Stutz}}]{Kainulainen13}
{Kainulainen}, J., {Ragan}, S.~E., {Henning}, T., \& {Stutz}, A. 2013, \aap,
  557, A120

\bibitem[{{Kainulainen} {et~al.}(2016{\natexlab{b}}){Kainulainen}, {Stutz},
  {Stanke}, {Abreu-Vicente}, {Beuther}, {Henning}, {Johnston}, \&
  {Megeath}}]{Kainulainen16b}
{Kainulainen}, J., {Stutz}, A.~M., {Stanke}, T., {et~al.} 2016{\natexlab{b}},
  ArXiv e-prints

\bibitem[{{K{\"o}nyves} {et~al.}(2010){K{\"o}nyves}, {Andr{\'e}},
  {Men'shchikov}, {Schneider}, {Arzoumanian}, {Bontemps}, {Attard}, {Motte},
  {Didelon}, {Maury}, {Abergel}, {Ali}, {Baluteau}, {Bernard}, {Cambr{\'e}sy},
  {Cox}, {di Francesco}, {di Giorgio}, {Griffin}, {Hargrave}, {Huang}, {Kirk},
  {Li}, {Martin}, {Minier}, {Molinari}, {Olofsson}, {Pezzuto}, {Russeil},
  {Roussel}, {Saraceno}, {Sauvage}, {Sibthorpe}, {Spinoglio}, {Testi},
  {Ward-Thompson}, {White}, {Wilson}, {Woodcraft}, \& {Zavagno}}]{Konyves10}
{K{\"o}nyves}, V., {Andr{\'e}}, P., {Men'shchikov}, A., {et~al.} 2010, \aap,
  518, L106

\bibitem[{{Krumholz} {et~al.}(2011){Krumholz}, {Leroy}, \&
  {McKee}}]{Krumholz11}
{Krumholz}, M.~R., {Leroy}, A.~K., \& {McKee}, C.~F. 2011, \apj, 731, 25

\bibitem[{{Li} {et~al.}(2014){Li}, {Esimbek}, {Zhou}, {Lou}, {Wu}, {Tang}, \&
  {He}}]{Li14}
{Li}, D.~L., {Esimbek}, J., {Zhou}, J.~J., {et~al.} 2014, \aap, 567, A10

\bibitem[{{Li} {et~al.}(2013){Li}, {Fang}, {Henning}, \& {Kainulainen}}]{Li13}
{Li}, H.-b., {Fang}, M., {Henning}, T., \& {Kainulainen}, J. 2013, \mnras, 436,
  3707

\bibitem[{{Malinen} {et~al.}(2012){Malinen}, {Juvela}, {Rawlings},
  {Ward-Thompson}, {Palmeirim}, \& {Andr{\'e}}}]{Malinen12}
{Malinen}, J., {Juvela}, M., {Rawlings}, M.~G., {et~al.} 2012, \aap, 544, A50

\bibitem[{{Marsh} {et~al.}(2015){Marsh}, {Whitworth}, \& {Lomax}}]{Marsh15}
{Marsh}, K.~A., {Whitworth}, A.~P., \& {Lomax}, O. 2015, \mnras, 454, 4282

\bibitem[{{Moeckel} \& {Burkert}(2015)}]{Moeckel15}
{Moeckel}, N. \& {Burkert}, A. 2015, \apj, 807, 67

\bibitem[{{Myers} {et~al.}(1983){Myers}, {Linke}, \& {Benson}}]{Myers83}
{Myers}, P.~C., {Linke}, R.~A., \& {Benson}, P.~J. 1983, \apj, 264, 517

\bibitem[{{Nishimura} {et~al.}(2015){Nishimura}, {Tokuda}, {Kimura}, {Muraoka},
  {Maezawa}, {Ogawa}, {Dobashi}, {Shimoikura}, {Mizuno}, {Fukui}, \&
  {Onishi}}]{Nishimura15}
{Nishimura}, A., {Tokuda}, K., {Kimura}, K., {et~al.} 2015, \apjs, 216, 18

\bibitem[{{Ossenkopf} \& {Henning}(1994)}]{Ossenkopf94}
{Ossenkopf}, V. \& {Henning}, T. 1994, \aap, 291, 943

\bibitem[{{Ostriker}(1964)}]{Ostriker64}
{Ostriker}, J. 1964, \apj, 140, 1056

\bibitem[{{Padoan} {et~al.}(2000){Padoan}, {Juvela}, {Bally}, \&
  {Nordlund}}]{Padoan00}
{Padoan}, P., {Juvela}, M., {Bally}, J., \& {Nordlund}, {\AA}. 2000, \apj, 529,
  259

\bibitem[{{Palmeirim} {et~al.}(2013){Palmeirim}, {Andr{\'e}}, {Kirk},
  {Ward-Thompson}, {Arzoumanian}, {K{\"o}nyves}, {Didelon}, {Schneider},
  {Benedettini}, {Bontemps}, {Di Francesco}, {Elia}, {Griffin}, {Hennemann},
  {Hill}, {Martin}, {Men'shchikov}, {Molinari}, {Motte}, {Nguyen Luong},
  {Nutter}, {Peretto}, {Pezzuto}, {Roy}, {Rygl}, {Spinoglio}, \&
  {White}}]{Palmeirim13}
{Palmeirim}, P., {Andr{\'e}}, P., {Kirk}, J., {et~al.} 2013, \aap, 550, A38

\bibitem[{{Panopoulou} {et~al.}(2017){Panopoulou}, {Psaradaki}, {Skalidis},
  {Tassis}, \& {Andrews}}]{Panopoulou17}
{Panopoulou}, G.~V., {Psaradaki}, I., {Skalidis}, R., {Tassis}, K., \&
  {Andrews}, J.~J. 2017, \mnras, 466, 2529

\bibitem[{{Panopoulou} {et~al.}(2014){Panopoulou}, {Tassis}, {Goldsmith}, \&
  {Heyer}}]{Panopoulou14}
{Panopoulou}, G.~V., {Tassis}, K., {Goldsmith}, P.~F., \& {Heyer}, M.~H. 2014,
  \mnras, 444, 2507

\bibitem[{{Peretto} {et~al.}(2012){Peretto}, {Andr{\'e}}, {K{\"o}nyves},
  {Schneider}, {Arzoumanian}, {Palmeirim}, {Didelon}, {Attard}, {Bernard}, {Di
  Francesco}, {Elia}, {Hennemann}, {Hill}, {Kirk}, {Men'shchikov}, {Motte},
  {Nguyen Luong}, {Roussel}, {Sousbie}, {Testi}, {Ward-Thompson}, {White}, \&
  {Zavagno}}]{Peretto12}
{Peretto}, N., {Andr{\'e}}, P., {K{\"o}nyves}, V., {et~al.} 2012, \aap, 541,
  A63

\bibitem[{{Pettitt} {et~al.}(2014){Pettitt}, {Dobbs}, {Acreman}, \&
  {Price}}]{Pettitt14}
{Pettitt}, A.~R., {Dobbs}, C.~L., {Acreman}, D.~M., \& {Price}, D.~J. 2014,
  \mnras, 444, 919

\bibitem[{{Pillai} {et~al.}(2015){Pillai}, {Kauffmann}, {Tan}, {Goldsmith},
  {Carey}, \& {Menten}}]{Pillai14}
{Pillai}, T., {Kauffmann}, J., {Tan}, J.~C., {et~al.} 2015, \apj, 799, 74

\bibitem[{{Planck Collaboration} {et~al.}(2016){Planck Collaboration}, {Adam},
  {Ade}, {Aghanim}, {Alves}, {Arnaud}, {Arzoumanian}, {Ashdown}, {Aumont},
  {Baccigalupi}, \& et~al.}]{Planck16}
{Planck Collaboration}, {Adam}, R., {Ade}, P.~A.~R., {et~al.} 2016, \aap, 586,
  A135

\bibitem[{{Pon} {et~al.}(2011){Pon}, {Johnstone}, \& {Heitsch}}]{Pon11}
{Pon}, A., {Johnstone}, D., \& {Heitsch}, F. 2011, \apj, 740, 88

\bibitem[{{Pon} {et~al.}(2012){Pon}, {Toal{\'a}}, {Johnstone},
  {V{\'a}zquez-Semadeni}, {Heitsch}, \& {G{\'o}mez}}]{Pon12}
{Pon}, A., {Toal{\'a}}, J.~A., {Johnstone}, D., {et~al.} 2012, \apj, 756, 145

\bibitem[{{S{\'a}nchez-Monge} {et~al.}(2014){S{\'a}nchez-Monge}, {Beltr{\'a}n},
  {Cesaroni}, {Etoka}, {Galli}, {Kumar}, {Moscadelli}, {Stanke}, {van der Tak},
  {Vig}, {Walmsley}, {Wang}, {Zinnecker}, {Elia}, {Molinari}, \&
  {Schisano}}]{Sanchez14}
{S{\'a}nchez-Monge}, {\'A}., {Beltr{\'a}n}, M.~T., {Cesaroni}, R., {et~al.}
  2014, \aap, 569, A11

\bibitem[{{Schneider} {et~al.}(2010){Schneider}, {Csengeri}, {Bontemps},
  {Motte}, {Simon}, {Hennebelle}, {Federrath}, \& {Klessen}}]{Schneider10}
{Schneider}, N., {Csengeri}, T., {Bontemps}, S., {et~al.} 2010, \aap, 520, A49

\bibitem[{{Sch{\"o}ier} {et~al.}(2005){Sch{\"o}ier}, {van der Tak}, {van
  Dishoeck}, \& {Black}}]{Schoier05}
{Sch{\"o}ier}, F.~L., {van der Tak}, F.~F.~S., {van Dishoeck}, E.~F., \&
  {Black}, J.~H. 2005, \aap, 432, 369

\bibitem[{{Seifried} {et~al.}(2016){Seifried}, {S{\'a}nchez-Monge}, {Walch}, \&
  {Banerjee}}]{Seifried16b}
{Seifried}, D., {S{\'a}nchez-Monge}, {\'A}., {Walch}, S., \& {Banerjee}, R.
  2016, \mnras, 459, 1892

\bibitem[{{Seifried} \& {Walch}(2015)}]{Seifried15}
{Seifried}, D. \& {Walch}, S. 2015, \mnras, 452, 2410

\bibitem[{{Seifried} \& {Walch}(2016)}]{Seifried16}
{Seifried}, D. \& {Walch}, S. 2016, \mnras, 459, L11

\bibitem[{{Shetty} {et~al.}(2009){Shetty}, {Kauffmann}, {Schnee}, {Goodman}, \&
  {Ercolano}}]{Shetty09}
{Shetty}, R., {Kauffmann}, J., {Schnee}, S., {Goodman}, A.~A., \& {Ercolano},
  B. 2009, \apj, 696, 2234

\bibitem[{{Smith} {et~al.}(2014){Smith}, {Glover}, \& {Klessen}}]{Smith14}
{Smith}, R.~J., {Glover}, S.~C.~O., \& {Klessen}, R.~S. 2014, \mnras, 445, 2900

\bibitem[{{Sugitani} {et~al.}(2011){Sugitani}, {Nakamura}, {Watanabe},
  {Tamura}, {Nishiyama}, {Nagayama}, {Kandori}, {Nagata}, {Sato}, {Gutermuth},
  {Wilson}, \& {Kawabe}}]{Sugitani11}
{Sugitani}, K., {Nakamura}, F., {Watanabe}, M., {et~al.} 2011, \apj, 734, 63

\bibitem[{{Sz{\H u}cs} {et~al.}(2014){Sz{\H u}cs}, {Glover}, \&
  {Klessen}}]{Szucs14}
{Sz{\H u}cs}, L., {Glover}, S.~C.~O., \& {Klessen}, R.~S. 2014, \mnras, 445,
  4055

\bibitem[{{Sz{\H u}cs} {et~al.}(2016){Sz{\H u}cs}, {Glover}, \&
  {Klessen}}]{Szucs16}
{Sz{\H u}cs}, L., {Glover}, S.~C.~O., \& {Klessen}, R.~S. 2016, \mnras, 460, 82

\bibitem[{{Vastel} {et~al.}(2006){Vastel}, {Caselli}, {Ceccarelli}, {Phillips},
  {Wiedner}, {Peng}, {Houde}, \& {Dominik}}]{Vastel06}
{Vastel}, C., {Caselli}, P., {Ceccarelli}, C., {et~al.} 2006, \apj, 645, 1198

\bibitem[{{Walch} {et~al.}(2015){Walch}, {Girichidis}, {Naab}, {Gatto},
  {Glover}, {W{\"u}nsch}, {Klessen}, {Clark}, {Peters}, {Derigs}, \&
  {Baczynski}}]{Walch15}
{Walch}, S., {Girichidis}, P., {Naab}, T., {et~al.} 2015, \mnras, 454, 238

\bibitem[{{Wilson}(1999)}]{Wilson99}
{Wilson}, T.~L. 1999, Reports on Progress in Physics, 62, 143

\end{thebibliography}
\end{document}